\documentclass[11pt,a4paper]{article}

\usepackage[margin=35mm]{geometry}
\usepackage[style=numeric, sorting=none, sortcites=true, bibstyle=ieee, citestyle=numeric-comp,, maxbibnames=99, giveninits=true]{biblatex}

\usepackage{authblk}

\usepackage{amssymb,amsmath,amsfonts,amsthm, amscd, mathrsfs, helvet, mathtools}
\usepackage{bm, stmaryrd}
\usepackage{graphicx, verbatim}
\usepackage[mathcal]{euscript}
\usepackage{color}
\usepackage[all]{xy}
\usepackage{relsize}

\usepackage{tikz}
\usetikzlibrary{cd}

\allowdisplaybreaks[1]

\usepackage{pict2e}

\usepackage{hyperref}
\hypersetup{
    colorlinks = true,
    linktocpage = true,
    linkcolor = blue,
    citecolor = blue,
    urlcolor = blue
}

\makeatletter
\DeclareFontFamily{OMX}{MnSymbolE}{}
\DeclareSymbolFont{MnLargeSymbols}{OMX}{MnSymbolE}{m}{n}
\SetSymbolFont{MnLargeSymbols}{bold}{OMX}{MnSymbolE}{b}{n}
\DeclareFontShape{OMX}{MnSymbolE}{m}{n}{
    <-6>  MnSymbolE5
   <6-7>  MnSymbolE6
   <7-8>  MnSymbolE7
   <8-9>  MnSymbolE8
   <9-10> MnSymbolE9
  <10-12> MnSymbolE10
  <12->   MnSymbolE12
}{}
\DeclareFontShape{OMX}{MnSymbolE}{b}{n}{
    <-6>  MnSymbolE-Bold5
   <6-7>  MnSymbolE-Bold6
   <7-8>  MnSymbolE-Bold7
   <8-9>  MnSymbolE-Bold8
   <9-10> MnSymbolE-Bold9
  <10-12> MnSymbolE-Bold10
  <12->   MnSymbolE-Bold12
}{}

\DeclareUnicodeCharacter{0301}{FIX ME!!!!}
\let\llangle\@undefined
\let\rrangle\@undefined
\DeclareMathDelimiter{\llangle}{\mathopen}%
                     {MnLargeSymbols}{'164}{MnLargeSymbols}{'164}
\DeclareMathDelimiter{\rrangle}{\mathclose}%
                     {MnLargeSymbols}{'171}{MnLargeSymbols}{'171}
\makeatother



\definecolor{myPurple}{rgb}{0.5,0.1,0.6}
\definecolor{myOrange}{rgb}{1.0,0.5,0.0}
\definecolor{myRed}{rgb}{1.0,0.0,0.0}
\definecolor{myGreen}{rgb}{0.0,0.5,0.0}
\definecolor{LatexBlue}{rgb}{0.211765,0.227451,0.666667}
\definecolor{myBlue}{rgb}{0.0,0.0,1.0}
\definecolor{myBlack}{rgb}{0.0,0.0,0.0}
\definecolor{myGray}{rgb}{0.3,0.3,0.3}

\theoremstyle{plain}
\newtheorem{theorem}{Theorem}[section]
\newtheorem*{theorem*}{Theorem}
\newtheorem{proposition}[theorem]{Proposition}
\newtheorem*{proposition*}{Proposition}
\newtheorem{lemma}[theorem]{Lemma}

\theoremstyle{definition}

\newtheorem{remarkx}[theorem]{Remark}

\newenvironment{remark}
  {\pushQED{\qed}\remarkx}
  {\popQED\endremarkx}

\newcommand{\tensor}[1]{{\mathfrak{#1}}}
\newcommand{\up}[1]{[#1]}
\newcommand{\uz}[1]{(#1)}

\DeclareMathOperator{\res}{res}

\DeclareMathOperator{\im}{im}

\DeclareMathOperator{\Ad}{Ad}

\newcommand{\SimTo}{\xrightarrow{\raisebox{-0.3 em}{\smash{\ensuremath{\text{\tiny$\cong$}}}}}}

\newcommand{\g}{\mathfrak{g}}
\newcommand{\C}{\mathcal{C}}
\newcommand{\CP}{\mathbb{P}^1}

\newcommand{\Lc}{\mathcal{L}}

\newcommand{\rd}{\mathrm{d}}

\newcommand{\Ec}{\mathcal{E}}

\newcommand{\dla}{\mathfrak{d}}%{\g^{[\bz]}}
\def\d{\mathfrak{d}}
\def\ft{\tilde{\mathfrak{f}}}
\def\g{\mathfrak{g}}

\def\id{\textup{id}}
\def\f{\mathfrak{f}}

\def\k{\mathfrak{k}}

\def\va{\mathfrak{v}}

\def\T{\mathcal{T}}

\def\C{\mathcal{C}}

\def\CS{\mathrm{CS}}

\def\jb{{\bm j}}
\def\rb{{\bm \rho}}
\def\jhb{\bm{\widehat{\jmath}}}

\newcommand{\bsb}[1]{\bm{[} #1 \bm{]}}
\newcommand{\brb}[1]{\bm{(} #1 \bm{)}}

\def\bzeta{{\bm \zeta}}
\def\bz{{\bm z}}
\def\bZ{\Pi {\bm z}}
\def\gzeta{\g^{\brb{\bzeta}}}

\newcommand{\lau}[1]{(\kern-.2em( #1 )\kern-.2em)}%{(( #1 ))}

\newcommand{\ms}[1]{\mathsf{#1}}

\def\CC{\mathbb{C}}
\def\CP{\mathbb{C}P^1}

\def\RR{\mathbb{R}}

\def\ZZ{\mathbb{Z}}

\def\ii{{\rm i}}

\def\1{\tensor{1}}
\def\2{\tensor{2}}
\def\3{\tensor{3}}
\def\4{\tensor{4}}

\numberwithin{equation}{section}

\date{}

\begin{document}

\title{\bf Integrable degenerate $\bm \Ec$-models\\
from 4d Chern-Simons theory}

\author[$\brb{\bzeta}$]{Joaquin Liniado}
\author[$\bsb{\bz}$]{Beno\^{\i}t Vicedo}

\affil[$\brb{\bzeta}$]{Instituto de Física La Plata (CONICET and Universidad Nacional de La Plata) CC 67 (1900) La Plata, Argentina}
\affil[$\bsb{\bz}$]{Department of Mathematics, University of York, York YO10 5DD, U.K.}

\maketitle

\begin{abstract}
We present a general construction of integrable degenerate $\Ec$-models on a $2$d manifold $\Sigma$ using the formalism of Costello and Yamazaki based on $4$d Chern-Simons theory on $\Sigma \times \CP$. We begin with a physically motivated review of the mathematical results of \cite{Benini:2020skc} where a unifying $2$d action was obtained from $4$d Chern-Simons theory which depends on a pair of $2$d fields $h$ and $\Lc$ on $\Sigma$ subject to a constraint and with $\Lc$ depending rationally on the complex coordinate on $\CP$. When the meromorphic $1$-form $\omega$ entering the action of $4$d Chern-Simons theory is required to have a double pole at infinity, the constraint between $h$ and $\Lc$ was solved in \cite{Lacroix:2020flf} to obtain integrable non-degenerate $\Ec$-models. We extend the latter approach to the most general setting of an arbitrary $1$-form $\omega$ and obtain integrable degenerate $\Ec$-models. To illustrate the procedure we reproduce two well known examples of integrable degenerate $\Ec$-models: the pseudo dual of the principal chiral model and the bi-Yang-Baxter $\sigma$-model.
\end{abstract}

\section{Introduction}

Given the action for a $2$d field theory, it is a difficult problem to decide whether or not it is classically integrable. Indeed, proving a theory is integrable requires showing, in particular, that the field equations of motion are equivalent to the flatness equation for a $2$d connection $\Lc$ valued in some finite-dimensional Lie algebra $\g$ and depending meromorphically on some auxiliary complex parameter. Unfortunately there is no systematic procedure for finding such a Lax connection when it exists.

In recent years, however, there has been tremendous progress towards the problem of classifying $2$d integrable field theories. In particular, Costello and Yamazaki proposed a general approach \cite{Costello:2019tri}, based on earlier work in the context of integrable spin chains \cite{Costello:2013sla, Costello:2013zra, Costello:2017dso, Costello:2018gyb}, for constructing $2$d integrable field theories starting from $4$d Chern-Simons theory. This approach, being rooted in the Lagrangian formalism, provides an elegant way of computing the $2$d action for the integrable field theory as well as its Lax connection. It has been extensively applied to reproduce many existing $2$d integrable field theories and also construct a wide variety of new ones \cite{Delduc:2019whp, Bykov:2019vkf, Bykov:2020nal, Bykov:2020tao, Lacroix:2020flf, Costello:2020lpi, Fukushima:2020dcp, Fukushima:2020kta, Schmidtt:2019otc, Schmidtt:2020dbf, Tian:2020ryu, Tian:2020pub, Caudrelier:2020xtn, Fukushima:2020tqv, Derryberry:2021rne}. See also \cite{Benini:2020skc, Ashwinkumar:2020krt, Bittleston:2020hfv, Penna:2020uky, Bittleston:2022cmz, Khan:2022vrx, Costello:2021zcl, Stedman:2021wrw, Levin:2022dnq, Zhou:2022mzb} for further developments in relation to $4$d Chern-Simons theory.

It is important to note that establishing the complete integrability of a $2$d field theory also requires showing that the integrals of motion constructed out of the Lax connection Poisson commute with one another, which necessitates moving to the Hamiltonian formalism. A general approach for constructing classical $2$d integrable field theories directly in the Hamiltonian formalism was proposed in \cite{Vicedo:2017cge}, and further developed in \cite{Delduc:2019bcl, Lacroix:2019xeh}, by starting from affine Gaudin models. It was shown in \cite{Vicedo:2019dej} by performing a Hamiltonian analysis of $4$d Chern-Simons theory (see also \cite{Schmidtt:2020dbf} for the $\ZZ_2$- and $\ZZ_4$-equivariant settings in the context of the $\lambda$-model) that the formalisms of \cite{Costello:2019tri} and \cite{Vicedo:2017cge} are intimately related. In particular, all the $2$d integrable field theories constructed from $4$d Chern-Simons theory are integrable in this stronger sense.

\medskip

The action of $4$d Chern-Simons theory is specified by a choice of meromorphic $1$-form $\omega$ on $\CP$ (one could also consider higher genus Riemann surfaces \cite{Costello:2019tri} but in this article we focus on the Riemann sphere). 
Different $2$d integrable field theories arise from different choices of $\omega$ and various other data to be reviewed in \S\ref{sec: review}.
The case when $\omega$ has at most double poles was studied in detail in \cite{Delduc:2019whp}, where a very simple `unifying' $2$d action was derived for all integrable field theories belonging to this class of meromorphic $1$-forms. A generalisation of this `unifying' $2$d action for arbitrary $\omega$ was then obtained in \cite{Benini:2020skc}, where a different perspective on the passage from $4$d Chern-Simons theory to $2$d integrable field theory was also advocated.

In order to write the `unifying' $2$d actions of \cite{Delduc:2019whp} or \cite{Benini:2020skc} in terms of the field of the $2$d integrable field theory alone, one needs to solve a certain constraint relating this field to the Lax connection. A very general class of solutions to this constraint was constructed in \cite{Lacroix:2020flf} when $\omega$ has arbitrary poles and zeroes in the complex plane but a double pole at infinity. This technical assumption was required in order to fix some of the gauge invariance of $4$d Chern-Simons theory and as a result remove any constant term from the Lax connection, as we recall at the start of \S\ref{sec:reductionto2d}. It was also shown in \cite{Lacroix:2020flf} that all of the $2$d integrable field theories arising from such solutions of the constraint are described by \emph{integrable} non-degenerate $\Ec$-models.

\medskip

Non-degenerate $\Ec$-models were introduced by Klim\v{c}\'\i{}k and \v{S}evera in \cite{Klimcik:1995dy, Klimcik:1995ux, Klimcik:1996np} as $\sigma$-models providing a natural setting for describing a non-abelian generalisation of $T$-duality, known as Poisson-Lie $T$-duality. Even though the generic non-degenerate $\Ec$-model is not integrable, it turns out that many interesting examples were found to be integrable \cite{Klimcik:2015gba, Klimcik:2017ken, Klimcik:2020fhs, Hoare:2020mpv}. A simple condition was also formulated on the data of the non-degenerate $\Ec$-model which ensures it is integrable \cite{Severa:2017kcs}. Of course, the data of the integrable non-degenerate $\Ec$-models constructed in \cite{Lacroix:2020flf} all satisfy this condition.

A generalisation of non-degenerate $\Ec$-models, known as degenerate $\Ec$-models or dressing cosets, was also introduced by Klim\v{c}\'\i{}k and \v{S}evera in \cite{Klimcik:1996np}. Whereas the non-degenerate $\Ec$-model can be described as a $\sigma$-model on a certain quotient space $K \setminus D$ where $K$ is an maximal isotropic subgroup of a Lie group $D$, the degenerate $\Ec$-model is a $\sigma$-model on a double quotient $K \setminus D \,/\, F$ where $F$ is another isotropic subgroup of $D$. It was recently shown in \cite{Klimcik:2019kkf} that a certain integrable $\sigma$-model that was constructed in \cite{Delduc:2017fib} provides an example of an \emph{integrable} degenerate $\Ec$-model. Very recently, conditions on the degenerate $\Ec$-model data ensuring its integrability, in the Hamiltonian sense recalled above, were also given in \cite{Klimcik:2021bqm} and a family of new integrable degenerate $\Ec$-models were constructed, including the pseudo dual chiral model and its multifield generalisations.

The purpose of this article is to extend the construction of \cite{Lacroix:2020flf} to the most general setting of an arbitrary meromorphic $1$-form $\omega$. In particular, we drop the technical assumption made in \cite{Lacroix:2020flf} that $\omega$ is required to have a double pole at infinity. We show that the solutions of the constraint equation from \cite{Benini:2020skc} which we construct by generalising the approach of \cite{Lacroix:2020flf} all give rise to \emph{integrable} degenerate $\Ec$-models. Just as in \cite{Lacroix:2020flf}, the Lie group $D$ is determined by the pole structure of $\omega$ and the maximal isotropic subgroup $K \subset D$ is determined by the choice of boundary condition imposed on the Chern-Simons field at the collection of poles of $\omega$. On the other hand, the isotropic subgroup $F$, which is specific to the present case, is a remnant of the gauge symmetry of the $4$d Chern-Simons theory under the Lie group $G$ and is given by the image of the diagonal embedding $G \hookrightarrow D$.

\medskip

The plan of the paper is as follows.

In \S\ref{sec: review} we review the alternative, less conventional approach of \cite{Benini:2020skc} for extracting the action of a $2$d integrable field theory from the $4$d Chern-Simons theory of Costello and Yamazaki \cite{Costello:2019tri}. One advantage of this approach is that it makes the passage from $4$d to $2$d more direct, with the field $h$ of the $2$d theory being introduced along surface defects in the $4$d theory to ensure gauge invariance.

In \S\ref{sec:reductionto2d} we generalise the approach of \cite{Lacroix:2020flf} for solving a constraint between the $2$d field $h$ and the $4$d gauge field appearing in the construction of \cite{Benini:2020skc}. More precisely, we do away with the technical requirement in \cite{Lacroix:2020flf} that the $1$-form $\omega$ in the $4$d Chern-Simons action should have a double pole at infinity. Starting from a general meromorphic $1$-form $\omega$, the resulting $2$d integrable field theories are degenerate $\Ec$-models.

In \S\ref{sec: examples} we give two detailed examples of the construction from \S\ref{sec:reductionto2d}. Namely, we apply the general formalism to recover the pseudo dual of the principal chiral model, or pseudo-chiral model for short, of Zakharov and Mikhailov \cite{Zakharov:1973pp} and the bi-Yang-Baxter $\sigma$-model proposed by Klim\v{c}\'\i{}k in \cite{Klimcik:2008eq}.

\section{A review on 4d Chern-Simons and 2d IFT} \label{sec: review}

In this section we begin by reviewing the correspondence between $4$d Chern-Simons theory and $2$d integrable field theories, proposed by Costello and Yamazaki in \cite{Costello:2019tri}. We will, however, follow the approach advocated in \cite{Benini:2020skc} which puts special emphasis on the principle of gauge invariance. In this approach, the $4$d Chern-Simons field $A$ is coupled to additional degrees of freedom, the so-called \emph{edge modes}, living on certain surface defects. This is to ensure the full gauge invariance of the theory. The $2$d integrable field theory is then seen to emerge in a particular gauge by going partly on-shell. Although the main ideas of \cite{Benini:2020skc} are intrinsically physical, the constructions rely heavily on methods of homotopical analysis and the theory of groupoids. So the purpose of this section is to review the key steps of the approach of \cite{Benini:2020skc} using a language more familiar to theoretical physicists.

\subsection{4d Chern-Simons theory} \label{sec: 4d CS action}

Let $X \coloneqq \Sigma\times \mathbb{C}P^1$ where $\Sigma$ denotes a $2$d manifold which will eventually correspond to the space-time of the $2$d integrable field theory. We take $\Sigma=\mathbb{R}^2$ or $\mathbb{R}\times S^1$ with coordinates $(\tau,\sigma)$.

Let $G$ be a real, simply connected Lie group with Lie algebra $\g$. Let $\g^\CC \coloneqq \g \otimes_\RR \CC$ be the complexification of $\g$ and let $G^\CC$ denote the corresponding complex Lie group. We fix a non-degenerate, symmetric and ad-invariant bilinear form $\langle\cdot,\cdot\rangle:\g \times \g \to \RR$ and denote by $\langle\cdot,\cdot\rangle:\g^\CC \times \g^\CC \to \CC$ its complex linear extension to $\g^\CC$.

\subsubsection{The meromorphic 1-form} \label{sec: omega}

The key ingredient entering the definition of the $4$d variant of Chern-Simons theory \cite{Costello:2019tri} is a choice of meromorphic 1-form $\omega$ on $\mathbb{C}P^1$.

We denote by $\bZ \subset \mathbb{C}P^1$ the set of poles of $\omega$ and by $n_x$ the order of the pole $x \in \bZ$. Let $\bZ' \coloneqq \bZ \setminus \{\infty\}$ be the subset of finite poles of $\omega$. The reason for this notation will be justified shortly. Fixing a coordinate $z$ on $\mathbb{C} \subset \mathbb{C}P^1$ we can write $\omega$ explicitly as
\begin{equation} \label{omega def}
\omega=\left(\sum_{x\in \bZ' }\sum_{p=0}^{n_x-1}\frac{\ell^x_p}{(z-x)^{p+1}}-\sum_{p=1}^{n_{\infty}-1}\ell^{\infty}_p z^{p-1}\right)\rd z \eqqcolon \varphi(z)\rd z\,,
\end{equation}
for some $\ell^x_p\in \mathbb{C}$ which we call \emph{levels}. We impose reality conditions on each $x\in \bZ$ and its corresponding levels by requiring that $\overline{\varphi(z)} = \varphi(\bar{z})$. In particular, introducing the subset of real poles $\bz_{\mathrm{r}} \coloneqq \bz'_{\mathrm{r}} \,\sqcup\, \{\infty\}$, where $\bz'_{\mathrm{r}} \coloneqq \bZ' \,\cap\, \mathbb{R}$, the associated levels are real, i.e. $\ell^x_p \in \mathbb{R}$ for $x\in \bz_{\mathrm{r}}$. The remaining poles come in complex conjugate pairs and we define $\bz_{\mathrm{c}}\coloneqq \{x \in \bZ \,|\, \Im x > 0\}$ so that $\bZ = \bz_{\mathrm{r}}\,\sqcup \,\bz_{\mathrm c}\,\sqcup\, \bar{\bz}_{\mathrm c}$. For every $x \in \bz_{\mathrm{c}} \,\sqcup\, \bar{\bz}_{\mathrm c}$ we have $n_{\bar{x}} = n_x$ and $\overline{\ell^x_p}=\ell^{\bar{x}}_p$ for $p = 0,\dots, n_{x}-1$. It is convenient to introduce the subset $\bz \coloneqq  \bz_{\mathrm{r}} \,\sqcup\, \bz_{\mathrm{c}}$ of \emph{independent} poles, namely which are independent under complex conjugation. Finally, we introduce the subset $\bz'\coloneqq  \bz'_{\mathrm{r}}\,\sqcup\, \bz_{\mathrm{c}}\subset \bz$ of finite independent poles in $\bz$. We let $\Pi \coloneqq \{ \id, \ms t \}$ denote the group $\ZZ_2$ generated by the element $\ms t$ acting by complex conjugation on $\CP$. The notation $\bZ$ (resp. $\bZ'$) introduced above then corresponds to the orbit of the set $\bz$ (resp. $\bz'$) under $\Pi$.

We will also be interested in the set of zeroes of $\omega$ which can be similarly decomposed as $\bzeta_{\mathrm{r}} \, \sqcup \, \bzeta_{\mathrm{c}} \, \sqcup \, \bar{\bzeta}_{\mathrm{c}}$ with $\bzeta_{\mathrm{r}}\subset \mathbb{R}$ the subset of real zeroes and $\bzeta_{\mathrm{c}}\subset  \{y \in \mathbb{C} \,|\, \Im y > 0\}$ the subset of complex zeroes with positive imaginary part. Moreover, we introduce the
set $\bzeta \coloneqq  \bzeta_{\mathrm{r}} \,\sqcup \, \bzeta_{\mathrm{c}}$ of independent zeroes and let $m_y \in \mathbb{Z}_{\geq 1}$ denote the order of the zero $y \in \bzeta$. For $y \in \bzeta_{\mathrm{c}}$, $\omega$ also has a zero of order $m_{\bar{y}}\coloneqq  m_y$ at $\overline{y} \in \overline{\bzeta}_{\mathrm{c}}$. The set of all zeroes of $\omega$ is $\Pi \bzeta$ and the set of all finite zeroes is $\Pi \bzeta'$.

The total number of poles of $\omega$ (counting multiplicities) is related to the total number of zeroes of $\omega$ (counting multiplicities) by
\begin{equation} \label{poles vs zeroes}
\sum_{x \in \bZ} n_x = \sum_{y \in \Pi \bzeta} m_y + 2.
\end{equation}
We will assume that the total number of poles of $\omega$ (counting multiplicities) is even, so that the total number of zeroes of $\omega$ (counting multiplicities) also is by \eqref{poles vs zeroes}.

We summarize the notations for the different subsets of poles and zeroes of $\omega$ introduced above in the tables below.
\begin{center}
\begin{tabular}{||c | c||}
 \hline
 \multicolumn{2}{|c|}{Subsets of poles of $\omega$} \\ [0.5ex] 
 \hline\hline
 $\bZ$ & all \\
 \hline
 $\bZ'$ & finite \\ 
 \hline
 $\bz_{\mathrm{r}}$ & real \\
 \hline
 $\bz_{\mathrm{r}}'$ & finite and real \\
 \hline 
 $\bz_{\mathrm{c}}$ & positive imaginary part \\
 \hline
 $\bz$ & independent\\ 
 \hline
 $\bz'$ & finite and independent\\
 \hline
\end{tabular}
\qquad
\begin{tabular}{||c | c||}
 \hline
 \multicolumn{2}{|c|}{Subsets of zeroes of $\omega$} \\ [0.5ex] 
 \hline\hline
 $\Pi \bzeta$ & all \\
 \hline
 $\Pi \bzeta'$ & finite \\
 \hline
 $\bzeta_{\mathrm{r}}$ & real \\
 \hline
 $\bzeta'_{\mathrm{r}}$ & finite and real \\
 \hline 
 $\bzeta_{\mathrm{c}}$ & positive imaginary part \\
 \hline
 $\bzeta$ & independent\\ 
 \hline
 $\bzeta'$ & finite and independent\\ 
 \hline
\end{tabular}
\end{center}

Following \cite{Lacroix:2020flf}, to account for the fact that poles and zeroes of $\omega$ have multiplicities, it will sometimes be convenient to use the notation $\bsb{\bz}$ for the set of pairs $\up{x, p}$ with $x \in \bz$ and $p = 0, \ldots, n_x - 1$, and similarly $\brb{\bzeta}$ for the set of pairs $\uz{y, q}$ with $y \in \bzeta$ and $q=0,\ldots, m_y-1$. We will use other similar notations, such as $\bsb{\bZ}$ and $\brb{\Pi\bzeta}$ for the set of all poles and all zeroes of $\omega$ with multiplicities included, respectively.

Finally, we shall also often make use of the local coordinates $\xi_x \coloneqq z-x$ at any finite pole $x \in \bz'$ or finite zero $x \in \bzeta'$ and the local coordinate at infinity $\xi_\infty \coloneqq z^{-1}$ if $\infty \in \bz$ or $\infty \in \bzeta$. The expansion of the meromorphic $1$-form at each pole $x \in \bz$ can then be written uniformly as
\begin{equation} \label{omega expansion}
\iota_x \omega = \sum_{p=0}^{n_x-1} \ell^x_p \xi_x^{-p-1} d\xi_x.
\end{equation}
For the point at infinity we have used the fact that $z^{p-1} \rd z = - \xi_\infty^{-p-1} \rd \xi_\infty$ for any $p \in \ZZ$ and introduced the additional level $\ell^\infty_0 \coloneqq - \sum_{x\in \bZ'} \ell^x_0$ for convenience.

\subsubsection{The 4d action and space of fields} \label{sec: 4d action and field}

Given a choice of meromorphic $1$-form $\omega$ as described in \S\ref{sec: omega}, the $4$d Chern-Simons action for a $\g$-valued $1$-form $A$ on $X$ is given by \cite{Costello:2019tri}
\begin{equation} \label{ec:4dCSaction}
    S_{4d}(A)=\frac{i}{4\pi}\int_{X}\omega\wedge \CS(A),
\end{equation}
where $\CS(A) \coloneqq \langle A, \rd A + \tfrac{1}{3}[A,A]\rangle$ denotes the Chern-Simons $3$-form.

Strictly speaking, the action \eqref{ec:4dCSaction} only makes sense when $\omega$ has at most simple poles, i.e. when $n_x = 1$ for all $x \in \bz$. Indeed, when $\omega$ has higher order poles, i.e. $n_x > 1$ for some $x \in \bz$, the $4$-form $\omega \wedge \CS(A)$ is not locally integrable near the \emph{surface defects} $\Sigma_x \coloneqq \Sigma \times \{ x \}$ with $n_x > 1$ and therefore needs to be suitably regularised \cite[\S 3.1]{Benini:2020skc}. In what follows we will not need the precise form of the regularised action, only its variation under gauge transformations to be discussed shortly, so we refer to \cite[\S 3.1]{Benini:2020skc}, see also \cite{Li:2020ljm}, for details of this regularisation procedure. We only point out that the regularisation is `local' in the sense that it consists in modifying the $4$-form $\omega \wedge \CS(A)$ only locally in small neighbourhoods of the surface defects $\Sigma \times \{ x \}$ for each pole $x \in \bz$ of $\omega$. We will keep denoting the action as $S_{4d}(A)$ in the presence of higher order poles in $\omega$.

Note that the $\rd z$-component of $A$ drops out from the action \eqref{ec:4dCSaction} due to the presence of the meromorphic $1$-form $\omega = \varphi(z) \rd z$. Another way to say this is that \eqref{ec:4dCSaction} is trivially invariant under translations $A \mapsto A + \chi \rd z$ for any $\chi \in C^\infty(X, \g)$ and we can fix this invariance by simply focusing on gauge fields with no $\rd z$-component. This remains true when $\omega$ has higher order poles \cite{Benini:2020skc}. From now on we will therefore always focus on fields of the form
\begin{equation} \label{no dz component}
A = A_\tau \rd \tau + A_\sigma \rd \sigma + A_{\bar z} \rd \bar z.
\end{equation}

On the other hand, it is too restrictive to consider only smooth $\g$-valued $1$-forms $A$, namely \eqref{no dz component} where the components $A_\tau$, $A_\sigma$ and $A_{\bar z}$ are smooth $\g$-valued functions on $X$. Indeed, one can allow these component functions to be singular at the zeroes of $\omega$ provided that the Lagrangian $\omega \wedge \CS(A)$, or its regularised version in the case when $\omega$ has higher order poles, remains locally integrable there. More precisely, let us fix a partition $\bzeta = \bzeta_+ \sqcup \bzeta_-$ of the independent zeroes of $\omega$ such that $\sum_{y \in \bzeta_+} m_y = \sum_{y \in \bzeta_-} m_y$. In particular, if all the zeroes of $\omega$ are simple, which will be the case in all our examples, then the latter condition means $|\Pi \bzeta_+| = |\Pi \bzeta_-|$.
We will take the space of fields of $4$d Chern-Simons theory to consist of $\g$-valued $1$-forms as in \eqref{no dz component} such that:
\begin{itemize}
  \item[$(i)$] $A_\pm \coloneqq A_\tau \pm A_\sigma$ has singularities at $\Sigma \times \bzeta_\pm$ and is smooth elsewhere,
  \item[$(ii)$] $A_{\bar z}$ is smooth everywhere,
  \item[$(iii)$] the $4$-form $\omega \wedge \CS(A)$ is locally integrable near $\Sigma \times \bzeta$.
\end{itemize}
Condition $(iii)$ puts constraints on the type of singularities allowed in condition $(i)$ so that the action is well defined. On the other hand, condition $(ii)$ is consistent with the gauge fixing condition $A_{\bar z} = 0$ which we shall impose later on in \S\ref{sec: 2d IFT review} in order to describe integrable field theories, once we have established the gauge invariance of $4$d Chern-Simons theory in the next section.

\subsection{Gauge invariance}

Having defined the $4$d Chern-Simons action in \S\ref{sec: 4d CS action}, the next step in the approach of \cite{Benini:2020skc} is to study its gauge invariance. We therefore consider the variation of the $4$d Chern-Simons action under gauge transformations
\begin{equation} \label{gauge transf def}
A \longmapsto {}^g A \coloneqq gAg^{-1} - \rd g g^{-1}
\end{equation}
for $g \in C^{\infty}(X,G)$ an arbitrary smooth $G$-valued function. Notice once again, as in \S\ref{sec: 4d action and field}, that the $\rd z$-component of the term $\rd g g^{-1}$ will automatically drop out from the action due to the presence of the $1$-form $\omega$, so that the gauge transformation \eqref{gauge transf def} effectively acts on connections of the form \eqref{no dz component}.

In the case when $\omega$ has only simple poles, and the action takes the form \eqref{ec:4dCSaction}, we easily see that
\begin{align} \label{ec:gaugetransformation}
S_{4d}({}^g A) &= S_{4d}(A)+\frac{i}{4\pi}\int_X \omega\, \wedge\, \rd\langle g^{-1}\rd g,A\rangle \notag\\
&\qquad\qquad\qquad + \frac{i}{4\pi}\int_X \omega\, \wedge \,\langle g^{-1}\rd g,[g^{-1}\rd g,g^{-1}\rd g]\rangle \,.
\end{align}
The $4$d Chern-Simons action is thus manifestly \emph{not} gauge invariant. It is instructive to compare \eqref{ec:gaugetransformation} with the gauge variation of the usual $3$d Chern-Simons action. In particular, the first additional term generated on the right hand side of \eqref{ec:gaugetransformation} is not the integral of an exact differential precisely due to the presence of the $1$-form $\omega$. In fact, neither of the two additional terms in \eqref{ec:gaugetransformation} will vanish in general. Therefore, obtaining a better understanding of these two terms is key to being able to promote $4$d Chern-Simons theory to a gauge invariant theory.

Before stating the general result from \cite{Benini:2020skc}, it is helpful to first explain the result in the simplest case when $\omega$ has only simple poles. It can be shown, see \cite{Costello:2019tri, Delduc:2019whp}, that the first additional term on the right hand side of \eqref{ec:gaugetransformation} localises on the surface defects $\Sigma_x = \Sigma \times \{ x \}$. Explicitly, if we suppose for simplicity that all the poles are real, i.e. $\bz = \bz_{\rm r}$, then we have
\begin{equation*}
\frac{i}{4\pi}\int_X \omega\, \wedge\, \rd\langle  g^{-1}\rd g,A\rangle
= - \frac{1}{2} \sum_{x \in \bz} \ell^x_0 \int_{\Sigma_x} \langle g^{-1}\rd g, A \rangle|_{\Sigma_x}.
\end{equation*}
In turn, the right hand side of the above can be rewritten as a single integral over $\Sigma$ as follows. The collection $(A|_{\Sigma_x})_{x \in \bz}$ of the restrictions of $A \in \Omega^1(X, \g)$ to each surface defect $\Sigma_x$, $x \in \bz$ defines a $\g$-valued $1$-form on the disjoint union of surface defects $\sqcup_{x \in \bz} \Sigma_x$. Alternatively, this can also be thought of as defining a $1$-form on $\Sigma$ but valued in the direct product of Lie algebras $\d = \prod_{x \in \bz} \g$. Moreover, we have a map $\jb^\ast : \Omega^1(X,\g) \to \Omega^1(\Sigma, \d)$ given by $\jb^\ast A = (A|_{\Sigma_x})_{x \in \bz}$. Likewise, the collection $(g|_{\Sigma_x})_{x \in \bz}$ of the restrictions of $g \in C^\infty(X, G)$ to the surface defects defines a smooth function on $\Sigma$ valued in the direct product Lie group $D = \prod_{x \in \bz} G$, and we have a map $\jb^\ast : C^\infty(X,G) \to  C^\infty(\Sigma, D)$ given by $\jb^\ast g = (g|_{\Sigma_x})_{x \in \bz}$. Defining the bilinear form $\langle\!\langle \cdot, \cdot \rangle\!\rangle_{\d} : \d \times \d \to \RR$ as $\langle\!\langle (\ms u_x)_{x \in \bz}, (\ms v_x)_{x \in \bz} \rangle\!\rangle_{\d} = \sum_{x \in \bz} \ell^x_0 \langle \ms u_x, \ms v_x \rangle$ we may finally rewrite the first additional term on the right hand side of \eqref{ec:gaugetransformation} as
\begin{equation} \label{second term gt}
\frac{i}{4\pi}\int_X \omega\, \wedge\, \rd\langle  g^{-1}\rd g,A\rangle
= - \frac{1}{2} \int_{\Sigma} \langle\!\langle (\jb^\ast g)^{-1}\rd (\jb^\ast g), \jb^\ast A \rangle\!\rangle_{\d}.
\end{equation}
The second additional term in \eqref{ec:gaugetransformation} may similarly be rewritten as a WZ-term for an extension of the $D$-valued field $\jb^\ast g \in C^\infty(\Sigma, D)$ to $\Sigma \times [0,1]$, see \cite{Costello:2019tri, Delduc:2019whp}.

\subsubsection{Defect Lie algebra and Lie group} \label{sec: defect Lie alg}

When $\omega$ has higher order poles, it was shown in \cite{Benini:2020skc} that the above rewriting of the gauge variation \eqref{ec:gaugetransformation} of the $4$d Chern-Simons action goes through with the obvious modifications. In particular, instead of just restricting the $\g$-valued $1$-form $A$ on $X$ to each surface defect $\Sigma_x$ one should keep the first $n_x - 1$ orders in the Taylor expansion of $A$ near $\Sigma_x$. Correspondingly, the direct product Lie algebra $\d$ and Lie group $D$ need to be replaced by the defect Lie algebra and Lie group \cite{Benini:2020skc,Lacroix:2020flf}.

Let $\T_x \coloneqq \RR[\varepsilon_x]/ (\varepsilon_x^{n_x})$ for each real pole $x \in \bz_{\rm r}$ and $\T_x \coloneqq \CC[\varepsilon_x]/ (\varepsilon_x^{n_x})$ for each complex pole $x \in \bz_{\rm c}$. We define the \emph{defect Lie algebra} as the real Lie algebra
\begin{equation} \label{defect alg}
\d \coloneqq \prod_{x \in \bz_{\rm r}} \g \otimes_\RR \T_x \times \prod_{x \in \bz_{\rm c}} \g^\CC \otimes_\CC \T_x,
\end{equation}
where $\g^\CC \otimes_\CC \T_x$ is regarded as a Lie algebra over $\RR$.
The Lie algebra relations of $\d$ are given explicitly as
\begin{equation*}
\big[ \ms u \otimes \varepsilon^p_x, \ms v \otimes \varepsilon^q_y \big] = \delta_{xy} [\ms u, \ms v] \otimes \varepsilon^{p+q}_x
\end{equation*}
where $\varepsilon_x^{p+q}=0$ for $p+q\geq n_x$. The truncated polynomial Lie algebras $\g \otimes_\RR \T_x$ and $\g^\CC \otimes_\CC \T_x$ are sometimes referred to as Takiff algebras and we will refer to the integer $p$ as the \emph{Takiff} degree of the element $\ms u \otimes \varepsilon^p_x$. Note that
\begin{equation} \label{dim poles}
\dim \d = \dim \g \sum_{x \in \bz_{\rm r}} n_x + 2 \dim \g \sum_{x \in \bz_{\rm c}} n_x = \dim \g \sum_{x \in \bZ} n_x.
\end{equation}
Recall form \S\ref{sec: omega} that we are assuming the total number of poles of $\omega$ (counting multiplicities) to be even, which implies by \eqref{dim poles} that $\dim \d$ is even.

We define a non-degenerate invariant symmetric bilinear form
\begin{subequations} \label{form on gz rc}
\begin{equation}
\langle\!\langle \cdot, \cdot \rangle\!\rangle_{\d} : \d \times \d \longrightarrow \RR
\end{equation}
with respect to which all the factors in \eqref{defect alg} are orthogonal, for any $x,y \in \bz_{\rm r}$ we set
\begin{equation} \label{form on gz 0}
\big\langle{\mkern-4mu}\big\langle \ms u \otimes \varepsilon^p_x,\ms v \otimes \varepsilon^q_y \big\rangle{\mkern-4mu}\big\rangle_{\d} \coloneqq \delta_{xy} \, \ell^x_{p+q} \langle \ms u, \ms v \rangle,
\end{equation}
where $\ell^x_p = 0$ for all $p \geq n_x$, and for any $x, y \in \bz_{\rm c}$ we set
\begin{equation} \label{form on gz 1}
\big\langle{\mkern-4mu}\big\langle \ms u \otimes \varepsilon^p_x, \ms v \otimes \varepsilon^q_y \big\rangle{\mkern-4mu}\big\rangle_{\d} \coloneqq \delta_{xy} \, \big( \ell^x_{p+q} \langle \ms u, \ms v \rangle + \ell^{\bar x}_{p+q} \langle \tau \ms u, \tau \ms v \rangle \big).
\end{equation}
\end{subequations}
In the case when all poles are real and simple, i.e. $\bz = \bz_{\rm r}$ and $n_x = 1$ for all $x \in \bz$, the defect Lie algebra $\d$ reduces to the direct product Lie algebra $\prod_{x \in \bz} \g$ considered above in the motivating example.

An important subalgebra of the defect Lie algebra $\d$ which will play a central role in the description of \emph{degenerate} $\Ec$-models is the diagonal subalgebra. To introduce it, we define the diagonal map
\begin{equation} \label{diagonal map}
\Delta : \g \longrightarrow \g^{\times |\bz|} \subset \d, \qquad
\ms u \longmapsto (\ms u \otimes \varepsilon^0_x)_{x \in \bz}.
\end{equation}
Note that at complex points $x \in \bz_{\rm c}$ the defect Lie algebra contains a copy of the real Lie algebra $\g$ in Takiff degree $0$ so that $\g^{\oplus |\bz|}$ is indeed a subalgebra of $\d$. The image of \eqref{diagonal map} then defines the \emph{diagonal subalgebra} $\f \coloneqq \im \Delta \subset \d$.

One can also introduce a real Lie group with Lie algebra $\d$ which we will call the \emph{defect Lie group} and denote by $D$. As a set this is given by the direct product
\begin{equation}
\label{ec:defectliegroup}
D = \prod_{x \in \bz_{\rm r}} \big( G \times (\g \otimes_\RR \T'_x) \big) \times \prod_{x \in \bz_{\rm c}} \big( G^\CC \times (\g^\CC \otimes_\CC \T'_x) \big),
\end{equation}
where $\T'_x \coloneqq \varepsilon_x \RR[\varepsilon_x]/ (\varepsilon_x^{n_x})$ for $x \in \bz_{\rm r}$ and $\T'_x \coloneqq \varepsilon_x \CC[\varepsilon_x]/ (\varepsilon_x^{n_x})$ for $x \in \bz_{\rm c}$.
However, the general definition of the Lie group structure on $D$, which can be found in \cite{Vizman}, is quite involved so we will not include it here to avoid clutter. In practice, we will only require the group law on $D$ when discussing specific examples in \S\ref{sec: examples}, where the corresponding expressions will be explicitly stated. Finally, note that we have the diagonal embedding $\Delta : G \to G^{\times |\bz|} \subset D$ corresponding to \eqref{diagonal map} at the group level.

Just as in the above motivating example, the purpose of introducing the defect Lie algebra is that we then have a map \cite{Benini:2020skc,Lacroix:2020flf}
\begin{equation} \label{ec:jstaralg}
    \jb^*:\Omega^1(X,\g) \longrightarrow \Omega^1 (\Sigma, \d), \qquad
    A \longmapsto \bigg(\sum_{p=0}^{n_x-1}\frac{1}{p!} (\partial_{\xi_x}^p A)|_{\Sigma_x}\otimes \varepsilon_x^p\bigg)_{x\in \bz}
\end{equation}
where $(\partial_{\xi_x}^p A)|_{\Sigma_x} \in \Omega^{1}(\Sigma,\g)$ denotes the pullback of $\partial_{\xi_x}^p A$ to each surface defect $\Sigma_x$. In other words, $\jb^*$ sends a $\g$-valued $1$-form $A$ on $X$ to the first $n_x$ terms in its Taylor expansion at each surface defect $\Sigma_x$ for $x \in \bz$. Similarly, for smooth $G$-valued functions on $X$ we have a map
\begin{align} 
\label{ec:jstargroup}
\jb^*:C^{\infty}(X,G)&\longrightarrow C^{\infty}(\Sigma,D) \notag\\
g & \longmapsto \jb^*g=\bigg(g|_{\Sigma_x}, \sum_{p=1}^{n_x-1}\frac{1}{p!} \big(\partial_{\xi_x}^{p-1}(\partial_{\xi_x} g g^{-1}) \big)\big|_{\Sigma_x} \otimes \varepsilon_x^p\bigg)_{x\in \bz} \,.
\end{align}
This definition differs from the one given in \cite{Benini:2020skc} which applies to matrix Lie groups.

With the above definitions in place, we are now in a position to state one of the main results of \cite{Benini:2020skc}. Namely, for an arbitrary meromorphic $1$-form $\omega$ as in \eqref{omega def}, the variation \eqref{ec:gaugetransformation} of the (regularised) $4$d Chern-Simons action under an arbitrary gauge transformation \eqref{gauge transf def} can be expressed as
\begin{equation}
\label{ec:defectgaugetransf}
    S_{4d}({}^g A)=S_{4d}(A)-\frac{1}{2}\int_{\Sigma}\langle\!\langle (\jb^* g)^{-1}\rd (\jb^*g),\jb^* A \rangle \!\rangle_{\d}-\frac{1}{2}I_{\d}^{\rm WZ}[\jb^* g] \,,
\end{equation}
where we have introduced the standard WZ-term for a field $h\in C^{\infty}\left(\Sigma, D\right)$, namely
\begin{equation}
\label{ec:wesszumino}
  I_{\d}^{\rm WZ}[h]= -\frac{1}{6}\int_{\Sigma \times I}\langle\!\langle \widehat{h}^{-1}\rd \widehat{h},[\widehat{h}^{-1}\rd \widehat{h},\widehat{h}^{-1}\rd \widehat{h}]\rangle\!\rangle_{\d}
\end{equation}
where $I\coloneqq [0,1]$ and $\widehat{h}\in C^{\infty}\left(\Sigma\times I, D \right)$ is any smooth extension of $h$ to $\Sigma \times I$ with the property that $\widehat{h}=h$ near $\Sigma \times \{0\} \subset \Sigma \times I$ and $\widehat{h}=\mathrm{id}$ near $\Sigma \times \{1\} \subset \Sigma \times I$. Of course, the second term on the right hand side in \eqref{ec:defectgaugetransf} coincides with \eqref{second term gt} when $\omega$ has only simple poles. The virtue of the result \eqref{ec:defectgaugetransf} is that it holds for any meromorphic $1$-form $\omega$ with poles of arbitrary order.

\subsubsection{Isotropy and edge-modes}
\label{sec:isotropy}

As already anticipated, it is now clear from \eqref{ec:defectgaugetransf} that the 4d Chern-Simons action is not gauge invariant. However, gauge invariance of the theory may still be achieved upon imposing boundary conditions on both $\jb^*A$ and $\jb^*g$ at the surface defects, in order for the two additional terms appearing in \eqref{ec:defectgaugetransf} to vanish.

Recall that a Lie subalgebra $\k \subset \d$ is said to be \emph{isotropic} with respect to \eqref{form on gz rc} if $\langle\!\langle \ms x,\ms y \rangle\!\rangle_\d=0$ for every $\ms x, \ms y \in \k$. Given a subgroup $K\subset D$ whose Lie algebra $\k\subset \d$ is isotropic with respect to $\langle\!\langle \cdot,\cdot\rangle\!\rangle_\d$, we can impose the boundary conditions
\begin{equation} \label{strict BCs}
\jb^*A \in \Omega^{1}(\Sigma,\k) \qquad \text{and} \qquad \jb^*g \in C^{\infty}(\Sigma,K)
\end{equation}
on both the field $A$ and the gauge transformation parameter $g$, so that in particular $(\jb^* g)^{-1}\rd (\jb^*g)\in \Omega^{1}(\Sigma,\k)$. The action then becomes manifestly gauge invariant since the last two terms on the right hand side of \eqref{ec:defectgaugetransf} vanish due to isotropy.

There are, however, two important related issues with the boundary conditions in \eqref{strict BCs}. Firstly, the condition imposed on $A$ is a \emph{strict} boundary condition which equates $\jb^\ast A$, the restriction of $A$ to the surface defects, with a $\k$-valued gauge field on $\Sigma$. But in a gauge theory one should only compare gauge fields via gauge transformations and not via equalities. Secondly, the condition imposed on $g$ restricts the set of allowed gauge transformations, thereby partially breaking the gauge invariance we are trying to achieve. In particular, the strict boundary condition imposed on $A$ is preserved only by these restricted gauge transformations. Therefore, strictly speaking, even upon imposing boundary conditions, $4$d Chern-Simons theory is not a fully gauge invariant theory.

Now the role of gauge transformations is to identify physically indistinguishable field configurations, by killing \emph{would-be} degrees of freedom. So restricting the kind of gauge transformations we allow will resurrect some of these degrees of freedom from the dead \cite{Tong:2016kpv}. In particular, if we insist on establishing a fully gauge invariant theory then these \emph{resurrected} degrees of freedom must be included somehow. 

This brings us to the second main result of \cite{Benini:2020skc}. Both issues with the boundary condition \eqref{strict BCs} can be resolved by introducing a new degree of freedom living on the surface defects, namely a smooth $D$-valued field $h\in C^{\infty}(\Sigma,D)$ called the \emph{edge mode}. It was shown in \cite{Benini:2020skc} that $4$d Chern-Simons theory with the boundary conditions \eqref{strict BCs} is equivalent to $4$d Chern-Simons theory coupled to the edge mode by introducing the extended action
\begin{equation}
\label{ec:edgemodeaction}
    S_{4d}^{\rm ext}(A,h)=S_{4d}(A)-\frac{1}{2}\int_{\Sigma}\langle\!\langle h^{-1}\rd h,\jb^* A \rangle \!\rangle_\d - \frac{1}{2}I_{\d}^{\rm WZ}[h] \,,
\end{equation}
together with the alternate boundary condition 
\begin{equation}
\label{ec:edgemodecondition}
{}^h (\jb^*A) \in \Omega^1(\Sigma,\k)\,.    
\end{equation}
That is, instead of imposing boundary conditions on $A$ and $g$ as in \eqref{strict BCs}, we only impose the boundary condition on $A$ and only up to a gauge transformation by $h$.

One can verify, using \eqref{ec:defectgaugetransf}, the Polyakov-Wiegmann identity \cite{POLYAKOV1983121} and the invariance of the bilinear form, that both the extended action \eqref{ec:edgemodeaction} and the constraint \eqref{ec:edgemodecondition} are invariant under the gauge transformation
\begin{equation}
\label{ec:Gsymmetry}
    A\longmapsto {}^g A\,, \qquad h \longmapsto h (\jb^* g)^{-1}
\end{equation}
with arbitrary $g\in C^{\infty}(X,G)$. Thus, we have defined a fully gauge invariant theory, at the price of adding a new field.

Observe that if we restrict the edge mode $h$ to take values in $K$ then we recover the original $4$d Chern-Simons action together with the original boundary conditions \eqref{strict BCs}. More precisely, the extended action \eqref{ec:edgemodeaction} enjoys the additional symmetry
\begin{equation}
\label{ec:gaugek}
    h \longmapsto k h
\end{equation}
for arbitrary $k\in C^{\infty}(\Sigma,K)$. The invariance of \eqref{ec:edgemodeaction} under \eqref{ec:gaugek}, can be verified using the Polyakov-Wiegmann identity, the constraint \eqref{ec:edgemodecondition} and the isotropy of $\k$.
In other words, the degrees of freedom added can be described by a smooth field on $\Sigma$ valued in the quotient $K \setminus D$.

\subsection{2d integrable field theories} \label{sec: 2d IFT review}

Although $4$d Chern-Simons theory coupled to the edge mode as described above is equivalent to the original $4$d Chern-Simons theory with boundary conditions \eqref{strict BCs}, the advantage of the former is that it leads more naturally to $2$d integrable field theories. In particular, the field content of the latter will correspond precisely to the edge mode degrees of freedom living on the defect. Moreover, the Lax connection of the $2$d integrable field theory will come directly from the gauge field $A$ of the $4$d Chern-Simons theory in \eqref{no dz component}.

There are, however, two glaring issues with interpreting the gauge field \eqref{no dz component} as a Lax connection. The first is that $A$ has a component along the $\rd \bar z$ direction whereas a Lax connection should be a $1$-form along $\Sigma$. The second is that as it stands $A$ is not meromorphic in the $z$-coordinate which we would like to interpret as the spectral parameter of the Lax connection.

The first issue is easily resolved. Indeed, we can partially fix the gauge invariance of \eqref{ec:edgemodeaction} using the gauge fixing condition $A_{\bar z} = 0$ in order to get rid of the undesired $\rd \bar{z}$-component. This is analogous to the axial gauge in electrodynamics and Yang-Mills theories, where one of the component of the gauge field is set to vanish.
We will suggestively denote the gauge field $A$ in this gauge by the letter $\mathcal{L}$. Note that there is a residual gauge symmetry \eqref{gauge transf def} by $g\in C^{\infty}(X,G)$ satisfying $\partial_{\bar{z}}g g^{-1}=0$.

The second issue is more problematic. In order to resolve it we will have to go partly on-shell, which we turn to next.

\subsubsection{Solving the bulk equations of motion} \label{sec: solving bulk eom}

The action \eqref{ec:edgemodeaction} defines a $4$-dimensional theory due to the presence of the `bulk' term $S_{4d}(\Lc)$ for the `bulk' field $\Lc = \Lc_\tau \rd \tau + \Lc_\sigma \rd \sigma$. To obtain a $2$-dimensional theory we will therefore restrict to solutions of the bulk equations of motion. Specifically, varying the action \eqref{ec:edgemodeaction} with respect to both $\Lc$ and $h$, subject to the constraint \eqref{ec:edgemodecondition}, we find the bulk and boundary field equations of motion \cite{Benini:2020skc}
\begin{subequations}
\begin{align}
\label{ec:bulkeom} \partial_{\bar z} \mathcal{L} &= 0 \quad \text{on} \quad \Sigma \times (\CP \setminus \bzeta),\\
\label{ec:boundaryeom} \rd_{\Sigma}(\jb^*\Lc)+\tfrac{1}{2}\left[\jb^*\Lc,\jb^*\Lc\right] &= 0 \quad \text{on} \quad \Sigma,
\end{align}
\end{subequations}
where $\rd_\Sigma$ denotes the de Rham differential on $\Sigma$.

The bulk equation of motion \eqref{ec:bulkeom} expresses the fact that the components of $\Lc = \Lc_\tau \rd \tau + \Lc_\sigma \rd \sigma$ are holomorphic along $\CP$ away from the set $\bzeta$ of zeroes of $\omega$. However, recall from condition $(i)$ in \S\ref{sec: 4d action and field} that we allow the light-cone components $\Lc_\pm = \Lc_\tau \pm \Lc_\sigma$ of the gauge field to have singularities at the subset of poles $\bzeta_\pm$, so long as the Lagrangian $\omega \wedge \CS(\Lc)$ remained integrable along each surface $\Sigma \times \{ y \}$ for $y \in \bzeta$. We can therefore take the components of $\Lc$ to be of the form
\begin{subequations} \label{L rational form}
\begin{equation} \label{ec:polestructureadm}
    \Lc_\mu = \sum_{y \in \bzeta'}\sum_{q=0}^{m_y-1}\frac{\Lc_\mu^{(y,q)}}{(z-y)^{q+1}}+\sum_{q=0}^{m_\infty-1}\Lc_\mu^{(\infty,q)}z^{q+1}+\Lc_\mu^{\rm c}
\end{equation}
for $\mu=\tau,\sigma$, where the coefficient functions $\Lc_\mu^{\rm c}\in C^{\infty}(\Sigma,\g)$, $\Lc_{\mu}^{\uz{y,q}} \in C^{\infty}(\Sigma,\g)$ for $\uz{y, q} \in \brb{\bzeta_{\rm r}}$ and $\Lc_{\mu}^{\uz{y,q}} \in C^{\infty}(\Sigma,\g^\CC)$ for $\uz{y, q} \in \brb{\bzeta_{\rm c}}$ are related by
\begin{equation} \label{E-model condition}
\Lc_\tau^{\uz{y,q}} = \epsilon_y \Lc_\sigma^{\uz{y,q}}
\end{equation}
\end{subequations}
with $\epsilon_y = \pm 1$ for $y \in \bzeta_\pm$. Note that \eqref{E-model condition} ensures that the light-cone component $\Lc_\pm$ only has poles in $\bzeta_\pm$, and not in $\bzeta_\mp$, as required by condition $(i)$ from \S\ref{sec: 4d action and field}.
To see why singularities of the form \eqref{L rational form} are allowed by condition $(iii)$, consider the case when $\omega$ has only simple poles so that the action takes the form \eqref{ec:4dCSaction}. The cubic term in the Chern-Simons $3$-form drops out since $\Lc$ only has legs along $\rd \sigma$ and $\rd \tau$ so that the bulk Lagrangian reads
\begin{subequations} \label{admissibility quadr cubic}
\begin{equation} \label{admissibility quadr}
\omega \wedge \CS(\Lc) = \omega \wedge \big( \langle \Lc_+, \bar \partial \Lc_- \rangle - \langle \Lc_-, \bar \partial \Lc_+ \rangle\big) \wedge \rd \sigma^+ \wedge \rd \sigma^-.
\end{equation}
It is then easy to see that this $4$-form is locally finite near the surface $\Sigma \times \{y\}$ for each $y \in \bzeta$. Notice, in particular, that the condition \eqref{E-model condition} is used here to guarantee that the set of poles of $\Lc_\pm$ are disjoint so that, for instance, the $\delta$-functions at the set $\bzeta_-$ arising from $\bar \partial \Lc_-$ are multiplied by poles in $\bzeta_+$ coming from $\Lc_+$. Moreover, note that the singularities of the form \eqref{ec:polestructureadm} are also consistent with the cubic term in the Lagrangian before moving to the gauge $A_{\bar z} = 0$, i.e. when $A_{\bar z}$ is non-zero but smooth as in condition $(ii)$ of \ref{sec: 4d action and field}, namely
\begin{equation} \label{admissibility cubic}
\omega \wedge \langle A, \tfrac{1}{3}[A,A]\rangle = -2 \omega \wedge \langle A_{\Bar{z}},[A_+,A_-]\rangle \wedge \rd \bar z \wedge \rd \sigma^+ \wedge \rd \sigma^-.
\end{equation}
\end{subequations}
The above remains true also when $\omega$ has higher order poles since the regularisation procedure, to make sense of the action in that case, only modifies the Lagrangian locally near the surface defects $\Sigma \times \{ x \}$ for each $x \in \bz$.

Gauge fields $\Lc$ of the form \eqref{ec:polestructureadm} satisfying the condition \eqref{E-model condition} were referred to in \cite{Benini:2020skc, Lacroix:2020flf} as being \emph{admissible}. More precisely, \eqref{E-model condition} is the most natural solution of the admissibility conditions given there, cf. \cite[Example 5.4]{Benini:2020skc} and \cite[\S 3.5 \& \S 4.2]{Lacroix:2020flf}. The observation we made above is that the admissibility condition can be traced back to the $4$d Chern-Simons Lagrangian as the requirement that it be locally integrable.
In fact, this new perspective on admissible solutions of the bulk equations of motion \eqref{ec:bulkeom} leads to the following observation which will be useful later.

\begin{remark} \label{rem: stronger poles in L}
The rational expression \eqref{ec:polestructureadm}, with poles of order $m_y$ at each $y \in \bzeta$, is not the most general one for which the bulk action is locally integrable. Indeed, one could take poles of order $m_y + 1$ at each $y \in \bzeta$ while maintaining the local integrability of the expression \eqref{admissibility quadr} and also of \eqref{admissibility cubic} before fixing the gauge $A_{\bar z} = 0$. This is because both of the top forms in \eqref{admissibility quadr cubic} are locally integrable near a \emph{simple} pole of the component \cite[Lemma 2.1]{Benini:2020skc}. Note that this is precisely why the $4$d Chern-Simons action \eqref{ec:4dCSaction} was well defined in the case when $\omega$ has only simple poles. The reason we have kept the strength of the poles in \eqref{ec:polestructureadm} as they are is to ensure that the $2$d action we end up with is integrable, as we will see shortly.
\end{remark}

Upon restricting the gauge field $\Lc$ to be a solution of the bulk equation of motion \eqref{ec:bulkeom} as in \eqref{ec:polestructureadm}, the bulk $4$-dimensional term in the action \eqref{ec:edgemodeaction} disappears and we are left with the $2$-dimensional action
\begin{equation}
\label{ec:edgemodele1}
    S_{2d}(\mathcal{L},h)=-\frac{1}{2}\int_{\Sigma}\langle\!\langle h^{-1}\rd h,\jb^* \mathcal{L} \rangle \!\rangle_\d-\frac{1}{2}I_\d^{\rm WZ}[h] \,.
\end{equation}
The equation of motion of this action is the boundary equation of motion \eqref{ec:boundaryeom}. It was shown in \cite[Proposition 5.6]{Benini:2020skc} that for admissible solutions of the bulk equations of motion, the flatness equation \eqref{ec:boundaryeom} for $\jb^*\Lc$ lifts to a flatness equation for $\Lc$ itself, namely
\begin{equation} \label{flatness of Lc}
    \rd \Lc +\tfrac{1}{2}[\Lc,\Lc]=0 \quad \text{on }\Sigma\,.
\end{equation}
Here we can make use of the observation in Remark \ref{rem: stronger poles in L} by noting that the argument in the proof of \cite[Proposition 5.6]{Benini:2020skc} still applies if we increase the order of \emph{one} of the poles of $\Lc$ with components \eqref{ec:polestructureadm} by $1$. In other words, although the requirement that the action be well defined allows us to increase the order of \emph{all} the poles in $\Lc$ by $1$, the requirement that the boundary equations of motion \eqref{ec:boundaryeom} lift to the flatness of $\Lc$ in \eqref{flatness of Lc}, which ultimately ensures integrability, only enables us to increase the order of \emph{one} of the poles in $\Lc$ by $1$ while keeping the strength of all the other poles the same. We will see another proof of this later in \S\ref{sec: j hat iso}.

Finally, recall that after removing the $\rd \bar z$-component of the gauge field the gauge symmetry \eqref{ec:Gsymmetry} was restricted to those $g \in C^\infty(X, G)$ such that $\partial_{\bar z} g g^{-1} = 0$. And to preserve the pole structure of the meromorphic gauge field $\Lc$ in \eqref{ec:polestructureadm} we can restrict to $g \in C^\infty(\Sigma, G)$ which are independent of the coordinate on $\CP$. It follows from \eqref{ec:jstargroup} that for such gauge transformation parameters we have $\jb^\ast g = \Delta(g)$, where recall that $\Delta : G \to G^{\times |\bz|} \subset D$ is the diagonal embedding.

\subsubsection{The 2d action} \label{sec: 2d action}

The $2$-dimensional action \eqref{ec:edgemodele1} which we obtained from \eqref{ec:edgemodeaction} by solving the bulk equations of motion in \S\ref{sec: solving bulk eom} should, of course, be supplemented by the boundary condition given in \eqref{ec:edgemodecondition}. In other words, the fields $\Lc$ and $h$ on which the action \eqref{ec:edgemodele1} depends are not independent but instead are related by the constraint
\begin{equation} \label{constraint L h}
{}^h (\jb^*\Lc) \in \Omega^{1}(\Sigma,\k).
\end{equation}
The final step for obtaining a $2$-dimensional integrable field theory is therefore to solve the constraint \eqref{constraint L h} to find an expression for $\Lc$ in terms of $h$.

Indeed, \emph{suppose} that we can find a unique solution $\Lc=\Lc(h)$ to the constraint \eqref{constraint L h}. In order to respect the gauge invariance \eqref{ec:Gsymmetry} and \eqref{ec:gaugek} we further assume that this solution is such that
\begin{equation}
\label{ec:equivariance}
    \jb^*\Lc \big( kh\Delta(g)^{-1} \big) = {}^{\Delta(g)} \big( \jb^*\Lc(h) \big)
\end{equation}
for every $g \in C^{\infty}(\Sigma, G)$ and $k\in C^{\infty}(\Sigma, K)$. Note that the existence and uniqueness of such a solution depends on the choice of Lagrangian subalgebra $\k \subset \d$. One of the main results of \cite{Lacroix:2020flf} was to explicitly construct such solutions. The resulting models were shown to coincide with integrable non-degenerate $\Ec$-models. In the remainder of this article we will generalise the construction of \cite{Lacroix:2020flf} to obtain a more general class of solutions to \eqref{constraint L h} leading to the class of integrable \emph{degenerate} $\Ec$-models.

Given any solution of the boundary condition \eqref{constraint L h} satisfying the equivariance property \eqref{ec:equivariance}, the action \eqref{ec:edgemodele1} reduces to a $2$-dimensional action for the edge mode field $h \in C^{\infty}(\Sigma,D)$ \emph{alone} given by
\begin{equation}
\label{ec:2daction}
S_{2d}(h)=-\frac{1}{2}\int_\Sigma \langle \!\langle h^{-1}\rd h, \jb^*\mathcal{L}(h)\rangle\!\rangle_\d-\frac{1}{2}I_\d^{\rm WZ}[h] \,.
\end{equation}
By virtue of the property \eqref{ec:equivariance}, this action is invariant under the transformations
\begin{equation}
\label{ec:gaugesymmetry2daction}
    h \longrightarrow kh\Delta(g)^{-1}
\end{equation}
for any $k\in C^{\infty}(\Sigma,K)$ and $g\in C^{\infty}(\Sigma,G)$. Moreover, the equations of motion \eqref{flatness of Lc} which arose from the boundary equations of motion of the original $4$d Chern-Simons action now read
\begin{equation} \label{flatness of Lh}
    \rd \Lc(h) + \tfrac{1}{2}[\Lc(h),\Lc(h)]=0\,.
\end{equation}
In other words, the $2$-dimensional action \eqref{ec:2daction} has an associated Lax connection $\mathcal{L}(h)$ and therefore describes a $2$-dimensional integrable field theory.

\section{Obtaining degenerate \texorpdfstring{$\Ec$}{E}-models} \label{sec:reductionto2d}

The purpose of this section is to complete the passage from $4$d Chern-Simons theory to $2$d integrable field theories. In particular, we will show how to obtain integrable degenerate $\Ec$-models. Since the details of this section are quite technical, the reader interested in applying the construction may wish, on first read, to skip to \S\ref{sec: examples} where we present various examples of the procedure in detail. They may then refer back to the present section for further details of the construction. Before presenting these details, and in order to facilitate the reading of this section, we begin by giving a brief outline of the main strategy.

\medskip

As recalled in \S\ref{sec: 2d action}, the very last step in the approach of \cite{Benini:2020skc} for passing from $4$d Chern-Simons theory to $2$d integrable field theories consists in finding a solution of the constraint equation \eqref{constraint L h} which satisfies the transformation property \eqref{ec:equivariance}. Such solutions were constructed in \cite{Lacroix:2020flf} under the assumption that $\omega$ has a double pole at infinity. This technical assumption was used to fix the gauge symmetry under $F$. Specifically, under the assumption that $n_\infty = 2$, the component of the edge mode $h \in C^\infty(\Sigma, D)$ at infinity is a field on $\Sigma$ valued in the semi-direct product $G \ltimes \g$. The latter can be brought to the identity by using the $F$ symmetry and the component of the $K$ symmetry associated with the point at infinity, see \cite[\S 3.6]{Lacroix:2020flf} for details. With the gauge symmetry under $F$ fixed in this way, the component at infinity of the constraint \eqref{constraint L h} forces the constant term in the Lax connection to vanish (note that since $\omega$ initially has a pole at infinity, the Lax connection cannot have poles there). Therefore, the Lax connections considered in \cite{Lacroix:2020flf} are of the special form
\begin{equation} \label{L from LV}
\Lc_\mu = \sum_{y \in \bzeta'}\sum_{q=0}^{m_y-1}\frac{\Lc_\mu^{(y,q)}}{(z-y)^{q+1}}.
\end{equation}
Moreover, the property \eqref{ec:equivariance} that the solution $\Lc = \Lc(h)$ is required to satisfy boils down to $\jb^*\Lc (kh) = \jb^*\Lc(h)$. And indeed, the solutions constructed in \cite{Lacroix:2020flf} were shown to have this property and the resulting $2$d integrable field theories were shown to coincide with integrable non-degenerate $\Ec$-models.

The main purpose of this section is to generalise the results of \cite{Lacroix:2020flf} to the case of a generic $1$-form $\omega$, as defined in \S\ref{sec: omega}. The key idea behind the approach of \cite{Lacroix:2020flf} for solving \eqref{constraint L h} is to construct an involution $\Ec : \d \SimTo \d$ on the defect Lie algebra with the property that
\begin{equation} \label{non-deg E-model condition}
\jb_\bz \Lc_\tau = \Ec (\jb_\bz \Lc_\sigma),
\end{equation}
where we wrote $\jb^\ast \Lc = \jb_z \Lc_\tau \rd \tau +\jb_z \Lc_\sigma \rd \sigma$ in components.
More precisely, the property \eqref{non-deg E-model condition} was the one imposed in \cite{Lacroix:2020flf} but it will have to be adapted in the present case, see \eqref{deg E-model condition} below.
The property \eqref{non-deg E-model condition} was then used in \cite{Lacroix:2020flf} as the starting point for solving the constraint \eqref{constraint L h}.

In order to build such an involution $\Ec : \d \SimTo \d$ satisfying \eqref{non-deg E-model condition}, observe that the relationship between $\Lc_\tau$ and $\Lc_\sigma$ is in fact very simple to describe in terms of the coefficients of these rational functions \eqref{L from LV}. Indeed, recall that these are related by \eqref{E-model condition}, namely $\Lc_\tau^{(y,q)} = \epsilon_y \Lc_\sigma^{(y,q)}$ where $\epsilon_y = \pm 1$ for each $y \in \bzeta$. The idea of \cite{Lacroix:2020flf} is then to build two isomorphisms
\begin{equation} \label{two isos}
\begin{tikzcd}
\d & R'_{\Pi \bzeta'}(\g^\CC)^\Pi \arrow[l, "\jb_\bz"', "\cong"] \arrow[r, "\bm \pi_{\bzeta'}", "\cong"'] & \g^{\brb{\bzeta'}}
\end{tikzcd}
\end{equation}
from a certain space of rational functions $R'_{\Pi \bzeta'}(\g^\CC)^\Pi$, where the components \eqref{L from LV} of the Lax connection live, to the defect Lie algebra $\d$ by Taylor expanding at each $x \in \bz$ and to a vector space $\g^{\brb{\bzeta'}}$ by extracting the coefficients at each pole $y \in \bzeta'$.

If we do not fix the gauge symmetry by $F$, as was done in \cite{Lacroix:2020flf}, then the components of the Lax connection still have a constant term compared to \eqref{L from LV} and can in general also have a pole at infinity, cf. \eqref{L rational form}. In this section we will adapt the construction of \cite{Lacroix:2020flf} summarised above to this case, in particular defining suitable generalisations of the above isomorphisms \eqref{two isos} in \S\ref{sec: pi zeta iso} and \S\ref{sec: j hat iso}. These will then be used in \S\ref{sec: E-operator} to build an involution $\Ec :\d \SimTo \d$ which is symmetric with respect to the bilinear form on $\d$ introduced in \S\ref{sec: defect Lie alg}. Finally, we will use the latter in \S\ref{sec: E-model} to construct solutions of the constraint \eqref{constraint L h} satisfying \eqref{ec:equivariance} and thereby obtain the action of integrable degenerate $\Ec$-models.

\subsection{The real vector space \texorpdfstring{$R_{\Pi\bzeta}(\g^\CC)^\Pi$}{Rg}} \label{sec: vs Rg}

Given a complex vector space $V$ we let $R_{\Pi\bzeta}(V)$ denote the space of $V$-valued rational functions with poles at each $y \in \Pi \bzeta$ of order at most $m_y$, the order of the zero $y$ of $\omega$. It will also be useful to define the subspace $R'_{\Pi\bzeta}(V) \subset R_{\Pi\bzeta}(V)$ of $V$-valued rational functions without constant term.

If $V$ is equipped with an anti-linear involution $\tau : V \to V$ then we can define an action of $\Pi$ on $V$ by letting $\ms t \in \Pi$ act as $\tau$. This then also lifts to an action of $\Pi$ on $R_{\Pi\bzeta}(V)$. We can also define an action of $\Pi$ on $R_{\Pi\bzeta}(V)$ by letting $\ms t \in \Pi$ act as the pullback by complex conjugation $\mu_{\ms t} : z \mapsto \bar z$. We let $R_{\Pi\bzeta}(V)^\Pi$ denote the real vector space of rational functions in $R_{\Pi\bzeta}(V)$ on which these two actions coincide. We will also make use of the subspace $R'_{\Pi\bzeta}(V)^\Pi \subset R_{\Pi\bzeta}(V)^\Pi$ of such rational functions without constant term.

In what follows we will either take $V = \g^\CC$ or $V = C^\infty(\Sigma, \g^\CC)$, where the action of $\Pi$ on the latter is induced from the action of $\Pi$ on $\g^\CC$.
Explicitly, an element $f \in R_{\Pi\bzeta}(\g^\CC)^\Pi$ is a $\Pi$-equivariant $\g$-valued rational function of the form
\begin{equation} \label{uz def}
f(z) = \sum_{y \in \Pi \bzeta'} \sum_{q=0}^{m_y-1} \frac{f^{\uz{y, q}}}{(z-y)^{q+1}} + \sum_{q=0}^{m_\infty-1} f^{(\infty, q)} z^{q+1} + f^{\rm c}
\end{equation}
where $f^{\rm c} \in \g$, $f^{\uz{y, q}} \in \g$ for all $\uz{y, q} \in \brb{\bzeta_{\rm r}}$ with $q = 0, \ldots, m_y - 1$, and $f^{\uz{y, q}} \in \g^\CC$ for all $\uz{y, q} \in \brb{\bzeta_{\rm c}}$ with $q = 0, \ldots, m_y - 1$ and $f^{\uz{\bar y, q}} = \tau f^{\uz{y, q}}$.

The dimension of the real vector space $R_{\Pi\bzeta}(\g^\CC)^\Pi$ is given by
\begin{equation} \label{dim Rg}
\dim R_{\Pi\bzeta}(\g^\CC)^\Pi = \dim \g \bigg( \sum_{y \in \Pi \bzeta} m_y + 1 \bigg).
\end{equation}
The term $\dim \g \sum_{y \in \Pi \bzeta} m_y$ comes from counting the degrees of freedom in the pole parts at each $y \in \Pi \bzeta$ of a generic rational function \eqref{uz def}, see in particular the second equality in \eqref{dim zeroes} later. The additional $\dim \g$ comes from the constant term $f^{\rm c}$ in \eqref{uz def}. An alternative way of counting the dimension of $R_{\Pi\bzeta}(\g^\CC)^\Pi$ is to consider instead the isomorphic space $R_{\Pi\bzeta}(\g^\CC)^\Pi \omega$ which consists of $\Pi$-equivariant $\g^\CC$-valued meromorphic $1$-forms with poles in $\bZ$. Its dimension is then given by
\begin{equation} \label{dim Rg 2}
\dim R_{\Pi\bzeta}(\g^\CC)^\Pi = \dim \g \bigg( \sum_{x \in \bZ} n_x - 1 \bigg),
\end{equation}
where the term $\dim \g \sum_{x \in \bZ} n_x$ comes from counting the degrees of freedom in the pole parts of $f \omega$ for $f \in R_{\Pi\bzeta}(\g^\CC)^\Pi$ at each $x \in \bZ$ and the additional $- \dim \g$ accounts for the fact that the sum of the residues of a meromorphic $1$-form vanishes.
Of course, the two expressions \eqref{dim Rg} and \eqref{dim Rg 2} coincide by virtue of \eqref{poles vs zeroes}.

A $\Pi$-equivariant $\g$-valued rational function $f \in R'_{\Pi\bzeta}(\g^\CC)^\Pi$ takes the form
\begin{equation} \label{upz def}
f(z) = \sum_{y \in \Pi \bzeta'} \sum_{q=0}^{m_y-1} \frac{f^{\uz{y, q}}}{(z-y)^{q+1}} + \sum_{q=0}^{m_\infty-1} f^{(\infty, q)} z^{q+1}.
\end{equation}
Note that the only difference with \eqref{uz def} is that the `constant' term $f^{\rm c} \in \g$ is missing in \eqref{upz def}. The constant rational functions in $R_{\Pi\bzeta}(\g^\CC)^\Pi$ form a subspace isomorphic to $\g$ and we have a direct sum decomposition
\begin{equation} \label{rat func split off constant}
R_{\Pi\bzeta}(\g^\CC)^\Pi = \g \dotplus R'_{\Pi\bzeta}(\g^\CC)^\Pi
\end{equation}
given explicitly by writing a function $f \in R_{\Pi \bzeta}(\g^\CC)^\Pi$ as in \eqref{uz def}, with the first two sums defining the component in $R'_{\Pi\bzeta}(\g^\CC)^\Pi$, cf. \eqref{upz def}, and the constant term $f^{\rm c} \in \g$ corresponding to the component in $\g$.

It is useful to adjoin another copy of $\g$ to the space $R_{\Pi\bzeta}(\g^\CC)^\Pi$ by considering the direct sum $R_{\Pi\bzeta}(\g^\CC)^\Pi \oplus \g$. It is important to note that this additional copy of $\g$ is distinct from the copy of $\g$ already present in \eqref{rat func split off constant}, representing the constant term in the rational function. It follows from comparing \eqref{dim poles} with \eqref{dim Rg 2} that
\begin{equation} \label{dim R plus g}
\dim \big( R_{\Pi\bzeta}(\g^\CC)^\Pi \oplus \g \big) = \dim \d.
\end{equation}

We define the symmetric bilinear form
\begin{subequations} \label{bilinear R zeta}
\begin{equation} \label{bilinear R zeta a}
\langle\!\langle \cdot, \cdot \rangle\!\rangle_\omega : \big( R_{\Pi\bzeta}(\g^\CC)^\Pi \oplus \g \big) \times \big( R_{\Pi\bzeta}(\g^\CC)^\Pi \oplus \g \big) \longrightarrow \RR,
\end{equation}
given for any $f, g \in R_{\Pi\bzeta}(\g^\CC)^\Pi$ and $\ms u, \ms v \in \g$ by
\begin{equation} \label{bilinear R zeta b}
\langle\!\langle (f, \ms u), (g, \ms v) \rangle\!\rangle_\omega \coloneqq \sum_{x \in \bZ} \res_x \langle f, g \rangle \omega + \langle \ms u, g^{\rm c} \rangle + \langle f^{\rm c}, \ms v \rangle.
\end{equation}
\end{subequations}
By a slight abuse of notation, we will often denote the restriction of \eqref{bilinear R zeta} to the subspace $R_{\Pi\bzeta}(\g^\CC)^\Pi$ also as $\langle\!\langle \cdot, \cdot \rangle\!\rangle_\omega$. That is, we will write $\langle\!\langle f, g \rangle\!\rangle_\omega \coloneqq \langle\!\langle (f, 0), (g, 0) \rangle\!\rangle_\omega$ for any $f, g \in R_{\Pi\bzeta}(\g^\CC)^\Pi$.

\begin{lemma}
The bilinear form \eqref{bilinear R zeta} is non-degenerate. Its restriction to $R_{\Pi\bzeta}(\g^\CC)^\Pi$ is degenerate and its restriction to $R'_{\Pi\bzeta}(\g^\CC)^\Pi$ is also non-degenerate.
\begin{proof}
Let us first show that the restriction to $R_{\Pi\bzeta}(\g^\CC)^\Pi$ is degenerate. Consider a constant function $f = f^{\rm c} \in R_{\Pi\bzeta}(\g^\CC)^\Pi$. Then for any $g \in R_{\Pi\bzeta}(\g^\CC)^\Pi$ the poles of the $1$-form $\langle f, g \rangle \omega$ are contained in $\bZ$ so it follows that $\langle\!\langle f, g \rangle\!\rangle_\omega = 0$.

Consider now the restriction of the bilinear form \eqref{bilinear R zeta} to $R'_{\Pi\bzeta}(\g^\CC)^\Pi$. We show that this is non-degenerate.
For every $f, g \in R'_{\Pi\bzeta}(\g^\CC)^\Pi$ we have
\begin{equation*}
\langle\!\langle f, g \rangle\!\rangle_\omega = \sum_{x \in \bZ} \res_x \langle f, g \rangle \omega.
\end{equation*}
Since $f$ is not constant, it has poles at some of the zeroes of $\omega$. By choosing $g$ to also have a pole at one of these same zeroes, we can ensure that some of the poles of the $1$-form $\langle f, g \rangle \omega$ lie outside the subset $\bZ$, namely in $\Pi \bzeta$. It is then possible to choose $g$ such that $\langle\!\langle f, g \rangle\!\rangle_\omega \neq 0$.

It is clear from the form of the additional two terms in \eqref{bilinear R zeta b} that the bilinear form on the whole of $R_{\Pi \bzeta}(\g^\CC)^\Pi \oplus \g$ is itself also non-degenerate.
\end{proof}
\end{lemma}

\subsection{The isomorphism \texorpdfstring{$\bm \pi_\bzeta : R_{\Pi\bzeta}(\g^\CC)^\Pi \SimTo \g \oplus \gzeta$}{pi zeta}} \label{sec: pi zeta iso}

A rational function $f \in R_{\Pi\bzeta}(\g^\CC)^\Pi$, as in \eqref{uz def}, is uniquely determined by the coefficients $f^{\uz{y, q}} \in \g$ for $\uz{y, q} \in \brb{\bzeta_{\rm r}}$ and $f^{\uz{y, q}} \in \g^\CC$ for $\uz{y, q} \in \brb{\bzeta_{\rm c}}$ at each of its real and complex poles, along with the constant term $f^{\rm c} \in \g$. It is therefore convenient to introduce the vector space in which the coefficients of such rational functions live.
Explicitly, we associate with the zeroes of $\omega$ the real vector space
\begin{equation} \label{gzeta def}
\gzeta \coloneqq \prod_{\uz{y, q} \in \brb{\bzeta_{\rm r}}} \g \times \prod_{\uz{y, q} \in \brb{\bzeta_{\rm c}}} \g^\CC
\end{equation}
where $\g^\CC$ is regarded as a real vector space. Its dimension is
\begin{equation} \label{dim zeroes}
\dim \gzeta = \dim \g \sum_{y \in \bzeta_{\rm r}} m_y + 2 \dim \g \sum_{y \in \bzeta_{\rm c}} m_y = \dim \g \sum_{y \in \Pi \bzeta} m_y.
\end{equation}

We now have an obvious isomorphism
\begin{align} \label{pi zeta def}
\bm \pi_\bzeta : R_{\Pi\bzeta}(\g^\CC)^\Pi &\overset{\cong}\longrightarrow \g \oplus \gzeta, \notag\\
f &\longmapsto \Big( f^{\rm c}, \big( f^{\uz{y, q}} \big)_{\uz{y, q} \in \brb{\bzeta}} \Big)
\end{align}
which takes a rational function in $R_{\Pi\bzeta}(\g^\CC)^\Pi$ and returns its constant term in $\g$ and the coefficients at each of its poles as an element of $\gzeta$.

We also extend this map to an isomorphism $\bm \pi_\bzeta : R_{\Pi\bzeta}(\g^\CC)^\Pi \oplus \g \SimTo \g \oplus \gzeta \oplus \g$ by letting it act trivially on the additional copy of $\g$ introduced in \S\ref{sec: vs Rg}.

We define the symmetric bilinear form, cf. \cite[(4.16)]{Lacroix:2020flf},
\begin{subequations} \label{bilinear g zeta}
\begin{equation} \label{bilinear g zeta a}
\langle\!\langle \cdot, \cdot \rangle\!\rangle_{\g \oplus \gzeta \oplus \g} : \big( \g \oplus \gzeta \oplus \g \big) \times \big( \g \oplus \gzeta \oplus \g \big) \longrightarrow \RR,
\end{equation}
given for any $\ms U = (\ms U^{\uz{y,q}})_{\uz{y,q} \in \brb{\bzeta}}, \ms V = (\ms V^{\uz{y,q}})_{\uz{y,q} \in \brb{\bzeta}} \in \gzeta$ and $\ms x, \ms x', \ms y, \ms y' \in \g$ by
\begin{align} \label{bilinear g zeta b}
&\langle\!\langle (\ms x, \ms U, \ms x'), (\ms y, \ms V, \ms y') \rangle\!\rangle_{\g \oplus \gzeta \oplus \g} \notag\\
&\qquad\qquad \coloneqq \sum_{y \in \bzeta} \sum_{\substack{p, q=0\\ p+q \geq m_y - 1}}^{m_y-1} \frac{2}{|\Pi_y|} \Re \bigg( \alpha_{p,q} \big\langle \ms U^{\uz{y,p}}, \ms V^{\uz{y,q}} \big\rangle \bigg) + \langle \ms x, \ms y' \rangle + \langle \ms x', \ms y \rangle,
\end{align}
\end{subequations}
where $\alpha_{p,q} \coloneqq - \frac{1}{(p+q+1-m_y)!} \partial_{\xi_y}^{p+q+1-m_y} \psi_y(\xi_y)$ for $p,q = 0, \ldots, m_y$ is symmetric under the exchange of $p$ and $q$. Here we wrote $\omega = \psi_y(\xi_y) \xi_y^{m_y} \rd \xi_y$ in the local coordinate $\xi_y$ at $y \in \bzeta$ where $\psi_y(\xi_y)|_y \neq 0$ using the fact that $\omega$ has a zero of order $m_y$ at $y$. We also denote by $\Pi_y \subseteq \Pi$ the stabiliser subgroup of $y$ under the action of $\Pi$ on $\CP$, and $|\Pi_y|$ is its order. Explicitly, we have $|\Pi_y| = 2$ for any real point $y \in \bzeta_{\rm r}$ and $|\Pi_y| = 1$ for any complex point $y \in \bzeta_{\rm c}$.

The following is an immediate generalisation of \cite[Lemma 4.3]{Lacroix:2020flf}.

\begin{lemma} \label{lem: pi zeta ip}
For any $f, g \in R_{\Pi \bzeta}(\g^\CC)^\Pi$ and $\ms u, \ms v \in \g$ we have
\begin{equation*}
\langle\!\langle \bm\pi_{\bzeta} (f, \ms u), \bm\pi_{\bzeta} (g, \ms v) \rangle\!\rangle_{\g \oplus \gzeta \oplus \g}
= \langle\!\langle (f, \ms u), (g, \ms v) \rangle\!\rangle_\omega.
\end{equation*}
\begin{proof}
The first term in the bilinear form \eqref{bilinear R zeta b} can be rewritten as
\begin{equation} \label{bilinear pairing zeros}
\langle\!\langle f, g \rangle\!\rangle_\omega = \sum_{x \in \bZ} \res_x \langle f, g \rangle \omega = - \sum_{y \in \Pi \bzeta} \res_y \langle f, g \rangle \omega = - \sum_{y \in \bzeta} \frac{2}{|\Pi_y|} \Re\big( \!\res_y \langle f, g \rangle \omega \big)
\end{equation}
where in the second equality we used the residue theorem and the fact that the poles of the meromorphic $1$-form $\langle f, g \rangle\omega$ belong to the set $\Pi \bz \sqcup \Pi \bzeta$.
The right hand side of \eqref{bilinear pairing zeros} can be evaluated more explicitly as follows
\begin{align*}
\langle\!\langle f, g \rangle\!\rangle_\omega &= - \sum_{y \in \bzeta} \sum_{p=0}^{m_y-1} \frac{2}{|\Pi_y|} \Re\Big( \!\res_y \big\langle f^{\uz{y,p}} \xi_y^{-p-1}, \psi_y(\xi_y) \xi_y^{m_y} g \big\rangle \rd \xi_y \Big)\\
&= - \sum_{y \in \bzeta} \sum_{p, q=0}^{m_y-1} \frac{2}{|\Pi_y|} \Re \bigg( \frac{1}{p!} \partial_{\xi_y}^p \big( \psi_y(\xi_y) \xi_y^{m_y-q-1} \big) \big|_y \big\langle f^{\uz{y,p}}, g^{\uz{y,q}} \big\rangle \bigg).
\end{align*}
In the first equality we used the fact that $g \omega$ is regular at $y \in \bzeta$ so that the only contribution to the residue is from the pole term $f^{\uz{y,p}} \xi_y^{-p-1}$ at $y$ in $f$ and we wrote $\omega$ locally in the coordinate $\xi_y$. In the second equality we took the residue and used the fact that the poles of $g$ at $x \neq y$ in the expression $\partial_{\xi_y}^p \big( \psi_y(\xi_y) \xi_y^{m_y} g \big) \big|_y$ with $p = 0, \ldots, m_y-1$ vanish since $\psi_y(\xi_y) \xi_y^{m_y}$ has a zero of order $m_y$ at $y$. Finally, we note that $\frac{1}{p!} \partial_{\xi_y}^p \big( \psi_y(\xi_y) \xi_y^{m_y-q-1} \big) \big|_y = \frac{1}{(p+q+1-m_y)!} \partial_{\xi_y}^{p+q+1-m_y} \psi_y(\xi_y)$ if $p+q \geq m_y - 1$ and is zero otherwise, from which the result now follows.
\end{proof}
\end{lemma}

\subsection{The isomorphism \texorpdfstring{$\jb_\bz : R_{\Pi\bzeta}(\g^\CC)^\Pi \SimTo \f^\perp$}{alpha}} \label{sec: j hat iso}

Let us introduce a linear map
\begin{align} \label{j z def}
\jb_{\bz} : R_{\Pi\bzeta}(\g^\CC)^\Pi &\longrightarrow \d, \notag\\
f &\longmapsto \bigg( \sum_{p=0}^{n_x - 1} \frac{1}{p!} \big( \partial^p_{\xi_x} f \big) \big|_x \otimes \varepsilon_x^p \bigg)_{x \in \bz},
\end{align}
which takes a rational function with poles at the zeroes of $\omega$ and returns the first $n_x$ terms in its Taylor expansion at each pole $x \in \bz$ of $\omega$ in the local coordinate $\xi_x$, where $\xi_x = z-x$ if $x \in \bz'$ and $\xi_\infty = z^{-1}$ if $\infty \in \bz$. The linear map \eqref{j z def} cannot be an isomorphism on dimensional grounds by \eqref{dim R plus g}. However, we will show below in Proposition \ref{prop: Cauchy-Vandermonde} that it is injective. Before doing so we will show that \eqref{j z def} maps the bilinear form \eqref{bilinear R zeta}, or rather its restriction to $R_{\Pi\bzeta}(\g^\CC)^\Pi$, to the bilinear form \eqref{form on gz rc} on $\d$.

\begin{lemma} \label{lem: bilinear forms 1}
For any $f, g \in R_{\Pi\bzeta}(\g^\CC)^\Pi$, we have $\langle\!\langle \jb_{\bz} f, \jb_{\bz} g \rangle\!\rangle_{\d} = \langle\!\langle f, g \rangle\!\rangle_\omega$.
\begin{proof}
Let $f, g \in R_{\Pi\bzeta}(\g^\CC)^\Pi$.
First note that by using the reality conditions at all of the poles of $\omega$ we may rewrite the bilinear form \eqref{bilinear R zeta b} more explicitly as a sum over the independent poles, namely
\begin{equation} \label{bilinear pairing poles}
\sum_{x \in \bZ} \res_x \langle f, g \rangle \omega = \sum_{x \in \bz} \frac{2}{|\Pi_x|} \Re \big( \!\res_x \langle f, g \rangle \omega \big). 
\end{equation}
Recall that $\Pi_x \subseteq \Pi$ denotes the stabiliser subgroup of $x \in \bz$.

Recall the explicit expression \eqref{omega def} for the meromorphic $1$-form, and in particular its expansion \eqref{omega expansion} at each pole $x \in \bz$. We then have
\begin{align*}
\langle\!\langle f, g \rangle\!\rangle_\omega &= \sum_{x \in \bz'} \sum_{p=0}^{n_x-1} \frac{2}{|\Pi_x|} \Re \bigg( \frac{\ell^x_p}{p!} \big( \partial_{\xi_x}^p \langle f, g \rangle \big)\big|_x \bigg)
+ \sum_{p=0}^{n_{\infty}-1} \frac{\ell^{\infty}_p}{p!} \big( \partial^p_{\xi_\infty}\langle f, g \rangle \big)\big|_\infty\\
&= \sum_{x \in \bz} \sum_{p=0}^{n_x-1} \sum_{q=0}^p \frac{2}{|\Pi_x|} \Re \bigg( \ell^x_p \bigg\langle \frac{1}{q!} (\partial_{\xi_x}^q f)|_x, \frac{1}{(p-q)!} (\partial_{\xi_x}^{p-q} g)|_x \bigg\rangle \bigg)\\
&= \sum_{x \in \bz} \sum_{q,r=0}^{n_x-1} \frac{2}{|\Pi_x|} \Re \bigg( \ell^x_{q+r} \bigg\langle \frac{1}{q!} (\partial_{\xi_x}^q f)|_x, \frac{1}{r!} (\partial_{\xi_x}^r g)|_x \bigg\rangle \bigg)
= \langle\!\langle \jb_{\bz} f, \jb_{\bz} g \rangle\!\rangle_{\d}
\end{align*}
where in the second last step we changed variable from $p$ to $r = p-q$ and used the convention that $\ell^x_p = 0$ for $p \geq n_x$. The last equality is by definition \eqref{form on gz rc} of the bilinear form on $\d$ and of the map $\jb_{\bz}$ in \eqref{j z def}.
\end{proof}
\end{lemma}

Any $\ms v \in \g$ defines a constant function on $R_{\Pi\bzeta}(\g^\CC)^\Pi$. Explicitly, in the notation of \eqref{uz def} we have $f^{\rm c} = \ms v$ and $f^{\uz{y, q}} = 0$ for every $\uz{y,q} \in \brb{\Pi \bzeta}$. By abuse of notation we will denote this rational function also as $\ms v$. Its image under $\jb_\bz$ is the element $\jb_\bz \ms v = \Delta \ms v \in \f$ of the diagonal subalgebra $\f = \Delta \g \subset \d$. Moreover, any element of $\f$ can be represented in this way.
Let
\begin{equation} \label{v def}
\va \coloneqq \jb_{\bm z}\big( R'_{\Pi\bzeta}(\g^\CC)^\Pi \big).
\end{equation}

\begin{proposition} \label{prop: Cauchy-Vandermonde}
The linear map $\jb_{\bz}$ in \eqref{j z def} is an isomorphism onto its image $\f^\perp$. In particular, we have the direct sum decomposition $\f^\perp = \f \dotplus \va$.
\begin{proof}
We will first show that $\jb_{\bz}\big( R_{\Pi\bzeta}(\g^\CC)^\Pi \big) \subset \f^\perp$.
To see this, let $f \in R_{\Pi\bzeta}(\g^\CC)^\Pi$ and $\Delta \ms v \in \f$ be arbitrary. Then by Lemma \ref{lem: bilinear forms 1} we have
\begin{equation*}
\langle\!\langle \jb_{\bz} f, \Delta \ms v \rangle\!\rangle_{\d} = \langle\!\langle \jb_{\bz} f, \jb_{\bz} \ms v \rangle\!\rangle_{\d} = \langle\!\langle f, \ms v \rangle\!\rangle_\omega = 0,
\end{equation*}
where the last step follows by the residue theorem since $\langle f, \ms v \rangle \omega$ is a meromorphic $1$-form with poles contained in the subset $\bZ$.

It remains to show that the map $\jb_\bz$ is injective. The result will then follow by virtue of \eqref{dim Rg 2} which can be rewritten as $\dim R_{\Pi\bzeta}(\g^\CC)^\Pi = \dim \d - \dim \f = \dim \f^\perp$ using \eqref{dim poles}. Equivalently, with the help of the bijection \eqref{pi zeta def} it is enough to show that $\jb_\bz \circ \bm \pi_\bzeta^{-1} : \g \oplus \gzeta \to \d$ is injective.
The coefficients of the expansions to order $n_x$ at all the poles $x \in \bZ$ are given by
\begin{equation*}
\frac{1}{p!} ( \partial_z^p f)|_x = \sum_{\uz{y,q} \in \brb{\Pi \bzeta'}} C^{\up{x, p}}_{\quad\;\; \uz{y,q}} f^{\uz{y,q}} + \sum_{q=-1}^{m_\infty - 1} C^{\up{x, p}}_{\quad\;\; \uz{\infty,q}} f^{(\infty, q)}
\end{equation*}
for all $\up{x,p} \in \bsb{\bZ}$, where we have incorporated the constant term $f^{(\infty, -1)} \coloneqq f^{\rm c}$ into the $(-1)^{\rm st}$ term of the second sum and introduced the coefficients
\begin{equation} \label{Cauchy mat}
C^{\up{x, p}}_{\quad\;\; \uz{y,q}} \coloneqq \binom{p+q}{p} \frac{(-1)^p}{(x-y)^{p+q+1}}, \qquad C^{\up{x, p}}_{\quad\;\; (\infty, q)} \coloneqq \binom {q+1} p x^{q+1-p}
\end{equation}
for all $\up{x,p} \in \bsb{\bZ}$ and $\uz{y,q} \in \brb{\Pi \bzeta}$ where $q = 0, \ldots, m_y - 1$ for all $y \in \bzeta \setminus \{ \infty \}$ and $q = -1, \ldots, m_\infty - 1$ for $y = \infty$.
The expressions in \eqref{Cauchy mat} are the components of what is known as a confluent Cauchy-Vandermonde matrix, see for instance \cite[Definition 13]{VAVRIN1997271}.
By combining \eqref{dim zeroes} and \eqref{poles vs zeroes} we find $\dim(\g \oplus \gzeta) = \dim \d - \dim \g$. The matrix specified by the components \eqref{Cauchy mat} is of dimension $\dim \d \times \dim(\g \oplus \gzeta)$ so removing $\dim \g$ columns (by removing the highest order term in the expansion at any of the poles $x \in \bZ$) we obtain a square confluent Cauchy-Vandermonde matrix. The result now follows form the fact that a \emph{square} confluent Cauchy-Vandermonde matrix is invertible \cite[Corollary 19]{VAVRIN1997271}.

The last part follows from applying the injective linear map $\jb_\bz$ to the direct sum decomposition \eqref{rat func split off constant}.
\end{proof}
\end{proposition}

Even though we will not explicitly need it in order to construct the action for the Lax connection of the degenerate $\Ec$-model in \S\ref{sec: E-model} below, it is useful to try to extend the injective linear map \eqref{j z def} to an isomorphism $R_{\Pi \bzeta}(\g^\CC)^\Pi \oplus \g \overset{\cong}\longrightarrow \d$. This will be useful in constructing the $\Ec$-operator $\Ec : \d \to \d$ on all of $\d$ in \S\ref{sec: E-operator} below. Explicitly, we want to construct an isomorphism
\begin{subequations} \label{alpha def}
\begin{equation} \label{alpha def map}
\jhb_\bz \coloneqq \jb_\bz \oplus \rb_\bz : R_{\Pi \bzeta}(\g^\CC)^\Pi \oplus \g \overset{\cong}\longrightarrow \d,
\end{equation}
where $\rb_\bz : \g \to \d$ is a linear map whose image is complementary to $\f^\perp$ in $\d$. We will now give a general procedure for constructing such a linear map by requiring that its image $\ft \coloneqq \im \rb_\bz$ be isotropic and perpendicular to $\va$, i.e. such that
\begin{equation} \label{alpha def prop 1}
\d = \f^\perp \dotplus \ft, \qquad \ft \subset (\ft \dotplus \va)^\perp, 
\end{equation}
and with the property that, for any $\ms x, \ms y \in \g$,
\begin{equation} \label{alpha def prop 2}
\langle\!\langle \rb_\bz \ms x, \Delta \ms y \rangle\!\rangle_{\d} = \langle \ms x, \ms y \rangle.
\end{equation}
\end{subequations}

For simplicity we will suppose in the following argument that $\omega$ does not have a pole at infinity. The construction could also be adapted to that case.
Recall that $m_\infty$ denotes the order of the zero of $\omega$ at infinity. We let $P : \CC \to \CC$ be a generic polynomial of order $m_\infty + 1$, which therefore contains $m_\infty + 2$ arbitrary coefficients in $\CC$. If $\ms x \in \g$ then $\ms x \, P$ is a $\g$-valued polynomial. Notice, however, that $\ms x \, P$ does \emph{not} lie in $R_{\Pi \bzeta}(\g^\CC)^\Pi$ since rational functions in $R_{\Pi \bzeta}(\g^\CC)^\Pi$, which are of the form \eqref{uz def}, contain a polynomial of order at most $m_\infty$ (and not $m_\infty + 1$). Nevertheless, by a slight abuse of notation we will write
\begin{equation*}
\jb_\bz (\ms x \, P) = \bigg( \sum_{p=0}^{n_x - 1} \frac{1}{p!} \big( \ms x (\partial^p_{\xi_x} P)|_x \big) \otimes \varepsilon_x^p \bigg)_{x \in \bz} \in \d
\end{equation*}
for the Taylor expansion of $\ms x \, P$ at each $x \in \bz$ up to order $n_x - 1$, cf. \eqref{j z def}. We then define the linear map
\begin{equation} \label{map rho}
\rb_\bz : \g \longrightarrow \d,\qquad
\ms x \longmapsto \jb_\bz (\ms x \, P).
\end{equation}
By construction $\ft = \im \rb_\bz$ is a complement to $\f^\perp = \im \jb_\bz$ in $\d$ since $\ms x \, P \not\in R_{\Pi \bzeta}(\g^\CC)^\Pi$. Next, we will argue how to fix the coefficients of the polynomial $P$ by imposing the second property in \eqref{alpha def prop 1} and the property \eqref{alpha def prop 2}.

We will first show how to ensure that $\jb_{\bz} (\ms x \, P) \in \va^\perp$. Suppose first that $m_\infty = 0$, i.e. that $\omega$ does not vanish at infinity. Then any $g \in R'_{\Pi \bzeta}(\g^\CC)^\Pi$, specifically \emph{without} constant term, has a zero of order at least $1$ at infinity so that the $1$-form $\langle \ms x \, P, g \rangle \omega$ is regular at infinity. In particular, all of its poles lie in $\bZ$. By the same computation as in the proof of Lemma \ref{lem: bilinear forms 1} one shows that $\langle\!\langle \jb_{\bz} (\ms x \, P), \jb_{\bz} g \rangle\!\rangle_{\d} = \langle\!\langle \ms x \, P, g \rangle\!\rangle_\omega = 0$ where the last step is by the residue theorem. Hence $\jb_{\bz} (\ms x \, P) \in \va^\perp$ where in this case $P$ is an arbitrary polynomial of degree $1$, i.e. with two arbitrary coefficients.

Suppose now that $m_\infty > 0$. Then we can write any $g \in R'_{\Pi \bzeta}(\g^\CC)^\Pi$ as $g = g' + g_\infty$ where $g_\infty$ is a $\g^\CC$-valued polynomial of order $m_\infty$ with no constant term, which therefore contains $m_\infty$ arbitrary coefficients, and $g'$ contains no polynomial term. Consider the equation $\langle\!\langle \jb_{\bz} (\ms x \, P), \jb_{\bz} g \rangle\!\rangle_{\d} = 0$. By the exact same reasoning as above we deduce $\langle\!\langle \jb_{\bz} (\ms x \, P), \jb_{\bz} g' \rangle\!\rangle_{\d} = 0$ whereas the condition $\langle\!\langle \jb_{\bz} (\ms x \, P), \jb_{\bz} g_\infty \rangle\!\rangle_{\d} = 0$ imposes a triangular system of $m_\infty$ linear equations on the coefficients of the polynomial $P$. By solving these equations we can then ensure that $\jb_{\bz} (\ms x \, P) \in \va^\perp$ where $P$ is a polynomial of degree $m_\infty + 1$ but with only two free coefficients.

The two remaining coefficients in the polynomial $P$ can be fixed by requiring that $\ft$ is isotropic, which amounts to imposing the condition $\langle\!\langle \rb_\bz \ms x, \rb_\bz \ms y \rangle\!\rangle_{\d} = 0$ for any $\ms x, \ms y \in \g$, and the additional property \eqref{alpha def prop 2}.

Note that the linear map \eqref{map rho} could be naturally incorporated into the linear map \eqref{j z def} by extending the space of rational functions $R_{\Pi \bzeta}(\g^\CC)^\Pi$ to include the polynomial functions of the form $\ms x \, P$ for $\ms x \in \g$. In other words, if we define the space $\widehat{R}_{\Pi \bzeta}(\g^\CC)^\Pi \cong R_{\Pi \bzeta}(\g^\CC)^\Pi \oplus \g$ consisting of rational functions of the form \eqref{uz def} but with the pole at infinity of order $m_\infty + 1$ rather than $m_\infty$ then the isomorphism \eqref{alpha def map} can equally be described as an isomorphism
\begin{equation*}
\jhb_\bz : \widehat{R}_{\Pi \bzeta}(\g^\CC)^\Pi \overset{\cong}\longrightarrow \d,
\end{equation*}
defined in exactly the same way as \eqref{j z def}, namely by Taylor expanding in the local coordinate $\xi_x$ at each $x \in \bz$ up to the order $n_x - 1$. This discussion relates back to the observation made in Remark \ref{rem: stronger poles in L} and in the paragraph after \eqref{flatness of Lc}, namely that the $\g$-valued gauge field $\Lc$, which after going on-shell was valued in $R_{\Pi \bzeta}(\g^\CC)^\Pi$, could in fact be taken to live in the larger space $\widehat{R}_{\Pi \bzeta}(\g^\CC)^\Pi$ just defined.

The key property of the isomorphism \eqref{alpha def map}, generalising that of Lemma \ref{lem: bilinear forms 1} and which we shall use to construct the $\Ec$-operator in the next section, is the following.

\begin{lemma} \label{lem: bilinear forms 2}
For any $f, g \in R_{\Pi\bzeta}(\g^\CC)^\Pi$ and $\ms u, \ms v \in \g$ we have
\begin{equation*}
\langle\!\langle \jhb_\bz (f, \ms u), \jhb_\bz (g, \ms v) \rangle\!\rangle_{\d} = \langle\!\langle (f, \ms u), (g, \ms v) \rangle\!\rangle_\omega.
\end{equation*}
\begin{proof}
We have
\begin{align*}
\langle\!\langle \jhb_\bz (f, \ms u), \jhb_\bz (g, \ms v) \rangle\!\rangle_{\d}
&= \langle\!\langle \jb_\bz f, \jb_\bz g \rangle\!\rangle_{\d} + \langle\!\langle \rb_\bz \ms u, \jb_\bz g \rangle\!\rangle_{\d}
+ \langle\!\langle \jb_\bz f, \rb_\bz \ms v \rangle\!\rangle_{\d}\\
&= \langle\!\langle f, g \rangle\!\rangle_{\omega} + \langle\!\langle \rb_\bz \ms u, \Delta g^{\rm c} \rangle\!\rangle_{\d}
+ \langle\!\langle \Delta f^{\rm c}, \rb_\bz \ms v \rangle\!\rangle_{\d}\\
&= \langle\!\langle f, g \rangle\!\rangle_{\omega} + \langle \ms u, g^{\rm c} \rangle
+ \langle f^{\rm c}, \ms v \rangle = \langle\!\langle (f, \ms u), (g, \ms v) \rangle\!\rangle_\omega.
\end{align*}
where in the first step we used the isotropy of $\ft$. In the second line we have used Lemma \ref{lem: bilinear forms 1} for the first term and the fact that $\ft \perp \va$ for the last two terms. The very last step is by definition \eqref{bilinear R zeta b} of the bilinear form on $R_{\Pi\bzeta}(\g^\CC)^\Pi \oplus \g$.
\end{proof}
\end{lemma}

\subsection{The \texorpdfstring{$\Ec$}{E}-operator} \label{sec: E-operator}

We now wish to define an operator $\Ec : \d \to \d$ which is symmetric with respect to the bilinear form \eqref{form on gz rc} on $\d$. To do so, we will first construct an operator \cite{Lacroix:2020flf}
\begin{equation} \label{Etilde def}
\widetilde{\Ec} : \g \oplus \gzeta \oplus \g \overset{\cong}\longrightarrow \g \oplus \gzeta \oplus \g
\end{equation}
which is symmetric with respect to the bilinear form \eqref{bilinear g zeta} on $\g \oplus \gzeta \oplus \g$. This may then be transferred to $\d$ using the isomorphism $\jhb_\bz$ in \eqref{alpha def map} and the isomorphism $\bm \pi_\bzeta : R_{\Pi \bzeta}(\g^\CC)^\Pi \oplus \g \SimTo \g \oplus \gzeta \oplus \g$ from \S\ref{sec: pi zeta iso}.

Recall from \S\ref{sec: 4d action and field} that we assumed we were given a partition $\bzeta = \bzeta_+ \sqcup \bzeta_-$ of the set of zeroes of $\omega$ such that $\sum_{y \in \bzeta_+} m_y = \sum_{y \in \bzeta_-} m_y$. The latter condition means, in particular, that the vector space $\gzeta$ introduced in \eqref{gzeta def} splits into a direct sum
\begin{equation} \label{gzeta decomp}
\gzeta = \g^{\brb{\bzeta_+}} \dotplus \g^{\brb{\bzeta_-}}
\end{equation}
of two subspaces with $\dim \g^{\brb{\bzeta_+}} = \dim \g^{\brb{\bzeta_-}}$. In \S\ref{sec: solving bulk eom} we then introduced signs $\epsilon_y = \pm 1$ for each $y \in \bzeta$ which we used to impose the condition \eqref{E-model condition} on the components of the gauge field. Given this data, we introduce the involution
\begin{align} \label{Etilde def explicit}
\widetilde{\Ec} : \g \oplus \gzeta \oplus \g &\overset{\cong}\longrightarrow \g \oplus \gzeta \oplus \g, \notag\\
\Big( f^{\rm c}, \big( f^{\uz{y,q}} \big)_{\uz{y,q} \in \brb{\bzeta}}, \ms u \Big) &\longmapsto \Big( \ms u, \big( \epsilon_y f^{\uz{y,q}} \big)_{\uz{y,q} \in \brb{\bzeta}}, f^{\rm c} \Big).
\end{align}
In other words, we leave the elements of $\g^{\brb{\bzeta_+}}$ fixed, we change the sign of those in $\g^{\brb{\bzeta_-}}$ and we flip the two additional copies of $\g$.
One easily checks that the involution \eqref{Etilde def explicit} is symmetric with respect to the bilinear form \eqref{bilinear g zeta}.

Following \cite{Lacroix:2020flf}, consider the isomorphism
\begin{equation} \label{C def}
\C \coloneqq \jb_\bz \bm \pi_{\bzeta}^{-1} : \g \oplus \gzeta \overset{\cong}\longrightarrow \f^\perp.
\end{equation}
We use this to transfer \eqref{Etilde def} to an operator on $\d$. Explicitly, we have the following.

\begin{lemma} \label{lem: E def}
The operator $\Ec \coloneqq \C \widetilde{\Ec} \C^{-1} : \d \overset{\cong}\longrightarrow \d$ is an involution and is symmetric with respect to the bilinear form \eqref{form on gz rc} on $\d$.
\begin{proof}
The involution property is immediate from that of $\widetilde{\Ec}$.

By combining Lemmas \ref{lem: pi zeta ip} and \ref{lem: bilinear forms 2} we have $\langle\!\langle \C \ms U, \C \ms V \rangle\!\rangle_\d = \langle\!\langle \ms U, \ms V \rangle\!\rangle_{\g \oplus \gzeta \oplus \g}$ for any $\ms U, \ms V \in \g \oplus \gzeta \oplus \g$. Therefore, for any $\ms x, \ms y \in \d$ we deduce
\begin{align*}
\langle\!\langle \ms x, \Ec \ms y \rangle\!\rangle_{\d}
&= \langle\!\langle \ms x, \C \widetilde{\Ec} \C^{-1} \ms y \rangle\!\rangle_{\d}
= \langle\!\langle \C^{-1} \ms x, \widetilde{\Ec} \C^{-1} \ms y \rangle\!\rangle_{\g \oplus \gzeta \oplus \g}\\
&= \langle\!\langle \widetilde{\Ec} \C^{-1} \ms x, \C^{-1} \ms y \rangle\!\rangle_{\g \oplus \gzeta \oplus \g}
= \langle\!\langle \C \widetilde{\Ec} \C^{-1} \ms x, \ms y \rangle\!\rangle_{\d}
= \langle\!\langle \Ec \ms x, \ms y \rangle\!\rangle_{\d}.
\end{align*}
Hence $\Ec$ is also symmetric, as required.
\end{proof}
\end{lemma}

\begin{lemma}
The restriction of the symmetric bilinear form $\langle\!\langle \cdot, \Ec \cdot \rangle\!\rangle_{\d} : \d \times \d \to \RR$ to $\f \subset \d$ is non-degenerate.
\begin{proof}
Let $\ms v \in \f$ and suppose that $\langle\!\langle \ms u, \Ec \ms v \rangle\!\rangle_{\d} = 0$ for all $\ms u \in \f$. Then $\Ec \ms v \in \f^\perp$ which is a contradiction since $\Ec \f = \ft$ is by definition a complement of $\f^\perp$ in $\d$.
\end{proof}
\end{lemma}

\subsection{Solving the constraint between \texorpdfstring{$\Lc$}{L} and \texorpdfstring{$h$}{h}} \label{sec: E-model}

Having introduced all of the necessary ingredients in the previous sections, we are finally in a position to complete the final step of the construction of $2$d integrable field theories from $4$d Chern-Simons theory.
Recall from \S\ref{sec: 2d action} that in order to write down the final $2$d action as in \eqref{ec:2daction} we need to have a solution $\Lc = \Lc(h)$ of the constraint \eqref{constraint L h} satisfying the transformation property \eqref{ec:equivariance}. We will now show how to construct solutions which give rise to integrable \emph{degenerate} $\Ec$-models.

Let $\va_\pm$ denote the eigenspaces of $\Ec$ restricted to $\va$ with eigenvalues $\pm 1$. These are the images of the subspaces $\g^{\brb{\bzeta_\pm}}$ under the isomorphism $\C$ defined in \eqref{C def}. Equivalently, we can describe $\va_\pm$ as the image of the spaces $R'_{\Pi\bzeta_\pm}(\g^\CC)^\Pi$ of $\Pi$-equivariant $\g^\CC$-valued rational functions with poles in $\Pi\bzeta_\pm$ (see the start of \S\ref{sec: vs Rg}), namely
\begin{equation} \label{v pm def}
\va_\pm \coloneqq \jb_{\bm z}\big( R'_{\Pi\bzeta_\pm}(\g^\CC)^\Pi \big).
\end{equation}
We have $\dim \va_+ = \dim \va_- = \tfrac 12 \dim \d - \dim \g$. Following \cite{Klimcik:2019kkf, Klimcik:2021bqm}, define the projection operators $W^\pm_h : \d \to \d$ by, see in particular \cite[(3.23)]{Klimcik:2021bqm},
\begin{equation} \label{projector W degenerate}
\ker W^\pm_h = \Ad_{h^{-1}} \k, \qquad
\im W^\pm_h = \f \oplus \va_\pm.
\end{equation}

The constraint \eqref{constraint L h} explicitly says $B_\pm \coloneqq \Ad_h (\jb_\bz \Lc_\pm) - \partial_\pm h h^{-1} \in C^\infty(\Sigma, \k)$, which can be rewritten as $\jb_\bz \Lc_\pm = \Ad_{h^{-1}} B_\pm + h^{-1} \partial_\pm h$. Applying the operator $W^\pm_h$ to both sides, and using the fact that $\Ad_{h^{-1}} B_\pm$ is valued in $\ker W^\pm_h$, we get
\begin{equation} \label{W on jL}
W^\pm_h(\jb_\bz \Lc_\pm) = W^\pm_h(h^{-1} \partial_\pm h).
\end{equation}
On the other hand, in terms of light-cone components the condition \eqref{E-model condition} is equivalent to the statement that $\Lc^{\uz{y,q}}_\pm = 0$ for any $y \in \bzeta_\mp$ and $q = 0, \ldots, m_y-1$. We can rewrite this in terms of the $\widetilde{\Ec}$-operator \eqref{Etilde def explicit} as $\widetilde{\Ec} \big( \bm \pi_\bzeta (\Lc_\pm -\Lc^{\rm c}_\pm) \big) = \pm \bm \pi_\bzeta (\Lc_\pm - \Lc^{\rm c}_\pm)$. Applying the isomorphism $\C$ from \eqref{C def} on both sides and using the definition of the $\Ec$-operator in Lemma \ref{lem: E def} we then obtain the equivalent condition
\begin{equation} \label{deg E-model condition}
\Ec ( \jb_\bz \Lc_\pm - \Delta \Lc^{\rm c}_\pm) = \pm ( \jb_\bz \Lc_\pm - \Delta \Lc^{\rm c}_\pm)
\end{equation}
which implies that $\jb_\bz \Lc_\pm - \Delta \Lc^{\rm c}_\pm \in \va_\pm$. Since $\Delta \Lc^{\rm c}_\pm \in \f$ it follows that $\jb_\bz \Lc_\pm \in \im W^\pm_h$ so that the left hand side of \eqref{W on jL} just becomes $\jb_\bz \Lc_\pm$. In other words, we arrive at the following solution
\begin{equation} \label{constraint solution W}
\jb_\bz \Lc_\pm(h) = W^\pm_h(h^{-1} \partial_\pm h)
\end{equation}
of the constraint \eqref{constraint L h}.

It remains to show that this solution satisfies the desired transformation property \eqref{ec:equivariance}. For this we need the following lemma.

\begin{lemma} \label{lem: properties of W}
For any $g \in C^\infty(\Sigma, G)$ and $k \in C^\infty(\Sigma, K)$ we have
\begin{equation*}
W^\pm_{kh\Delta(g)^{-1}} = \Ad_{\Delta(g)} \circ W^\pm_h \circ \Ad_{\Delta(g)^{-1}}.
\end{equation*}
\begin{proof}
The projection operator $W^\pm_{k h\Delta(g)^{-1}} : \d \to \d$ is defined by
\begin{equation*}
\ker W^\pm_{k h\Delta(g)^{-1}} = \Ad_{\Delta(g)} \Ad_{h^{-1}} \k, \qquad
\im W^\pm_{k h\Delta(g)^{-1}} = \f \oplus \va_\pm
\end{equation*}
where in the first equality we have used the fact that $\Ad_{k^{-1}} \k = \k$ since $k$ is valued in $K$.
The statement now follows using the fact that $\va$ is invariant under the adjoint action of the diagonal subgroup $F = \im \Delta$.
\end{proof}
\end{lemma}

The property \eqref{ec:equivariance} now easily follows using Lemma \ref{lem: properties of W}, namely we have
\begin{align*}
\jb_\bz \Lc_\pm(kh\Delta(g)^{-1}) &= W^\pm_{k h \Delta(g)^{-1}}\big( \Delta(g) h^{-1} k^{-1} \partial_\pm (k h \Delta(g)^{-1}) \big)\\
&= \Ad_{\Delta(g)} \circ W^\pm_h \big( \Ad_{h^{-1}} (k^{-1} \partial_\pm k) + h^{-1} \partial_\pm h - \Delta(g)^{-1} \partial_\pm \Delta(g) \big)\\
&= \Delta(g) \big( W^\pm_h ( h^{-1} \partial_\pm h ) \big) \Delta(g)^{-1} - \partial_\pm \Delta(g) \Delta(g)^{-1}\\
&= {}^{\Delta(g)} \big( \jb_\bz \Lc_\pm(h) \big),
\end{align*}
where in the third equality we have used the fact that $\Ad_{h^{-1}} (k^{-1} \partial_\pm k)$ is valued in $\Ad_{h^{-1}}\k = \ker W^\pm_h$ and $\Delta(g)^{-1} \partial_\pm \Delta(g)$ is valued in $\f$.

Substituting the solution \eqref{constraint solution W} into the $2$d action \eqref{ec:2daction} we get the degenerate $\Ec$-model action \cite[(3.22)]{Klimcik:2021bqm}
\begin{align} \label{E-model action}
S_{2d}(h) &= \frac{1}{2}\int \Big(\langle\!\langle h^{-1}\partial_-h,W^+_h (h^{-1}\partial_+h) \rangle\!\rangle \notag\\
&\qquad\qquad\qquad - \langle\!\langle h^{-1}\partial_+h,W^-_h (h^{-1}\partial_-h) \rangle\!\rangle\Big)\rd \sigma^{+}\wedge \rd \sigma^{-}-\frac{1}{2}I^{\mathrm{WZ}}[h] \,.
\end{align}
This model can be defined for more general Lie algebras $\d$ and $\Ec$-operators than the ones considered in this article. In general, the action \eqref{E-model action} does not describe a $2$d integrable field theory. However, in the present case where $\d$ is the defect Lie algebra \eqref{defect alg} and the $\Ec$-operator is as defined in \S\ref{sec: E-operator}, namely when the data originates from $4$d Chern-Simons theory as reviewed in \S\ref{sec: 2d IFT review}, the action \eqref{E-model action} describes an integrable field theory by construction. In particular, since the right hand side of \eqref{constraint solution W} takes values in $\f \oplus \va_\pm \subset \f^\perp$ and $\jb_\bz : R_{\Pi \bzeta}(\g^\CC)^\Pi \SimTo \f^\perp$ is an isomorphism by Proposition \ref{prop: Cauchy-Vandermonde}, we can apply its inverse to both sides to obtain the Lax connection. Namely, denoting this inverse by $\bm p : \f^\perp \SimTo R_{\Pi \bzeta}(\g^\CC)^\Pi$ we have
\begin{equation} \label{Lax connection E-model}
\Lc_\pm(h) = \bm p \big( W^\pm_h(h^{-1} \partial_\pm h) \big),
\end{equation}
which provides a Lax connection for the integrable degenerate $\Ec$-model \eqref{E-model action}.

Although it is immediate by construction that \eqref{Lax connection E-model} satisfies the flatness equation \eqref{flatness of Lh}, it is instructive to show this explicitly. We will need the following lemma which is an immediate generalisation of \cite[Lemma 4.6 \& Remark 4.7]{Lacroix:2020flf}. Recall the definition of the real vector spaces $\va_\pm$ in \eqref{v pm def}.

\begin{lemma} \label{lem: p property}
For any $\ms u_\pm \in \f \oplus \va_\pm$ we have $\bm p [\ms u_+, \ms u_-] = [\bm p \ms u_+, \bm p \ms u_-]$.
\begin{proof}
Let $\ms u_\pm \in \f \oplus \va_\pm$ which we can write as $\ms u_\pm = \jb_\bz f_\pm$ for some $f_\pm \in R_{\Pi\bzeta_\pm}(\g^\CC)^\Pi$. Then we have
\begin{align*}
[\bm p \ms u_+, \bm p \ms u_-] = [f_+, f_-] = \bm p \jb_\bz [f_+, f_-] = \bm p [\jb_\bz f_+, \jb_\bz f_-] = \bm p [\ms u_+, \ms u_-].
\end{align*}
To see the second step, first observe that $[f_+, f_-] \in R_{\Pi \bzeta}(\g^\CC)^\Pi$ by virtue of the fact that the poles of $f_+$, which lie in $\bzeta_+$, are disjoint from those of $f_-$, which lie in $\bzeta_-$. Therefore $\jb_\bz$ has a well defined action on $[f_+, f_-]$ and we can  insert the identity in the form $\id = \bm p \jb_\bz$ out front. In the third step we then used the fact that the linear map $\jb_\bz$, defined by taking the truncated Taylor expansions at the points $x \in \bz$ is in fact a morphism of Lie algebras.
\end{proof}
\end{lemma}

The equations of motion of the degenerate $\Ec$-model \eqref{E-model action} take the form of a flatness equation \cite[(2.8)]{Klimcik:2021bqm}
\begin{equation*}
\partial_+ \big( W^-_h(h^{-1} \partial_- h) \big) - \partial_- \big( W^+_h(h^{-1} \partial_+ h) \big) + \big[ W^+_h(h^{-1} \partial_+ h), W^-_h(h^{-1} \partial_- h) \big] = 0
\end{equation*}
in $\d$. Applying the linear map $\bm p : \f^\perp \SimTo R_{\Pi \bzeta}(\g^\CC)^\Pi$ to both sides and using Lemma \ref{lem: p property} on the commutator term we obtain the desired flatness equation \eqref{flatness of Lh} for the Lax connection with light-cone components \eqref{Lax connection E-model}.

\begin{remark}
The action of the non-degenerate $\Ec$-model \cite[(2.5)]{Klimcik:2021bqm} can be written in exactly the same form as in \eqref{E-model action} but where the projectors $W^\pm_h$ are now defined by the conditions \cite[(2.6)]{Klimcik:2021bqm}
\begin{equation*}
\ker W^\pm_h = \Ad_{h^{-1}} \k, \qquad
\im W^\pm_h = \va_\pm
\end{equation*}
instead of \eqref{projector W degenerate}. Of course, this action for the non-degenerate $\Ec$-model is equivalent to the one derived in \cite{Lacroix:2020flf} (see \cite[\S 2.2]{Lacroix:2020flf}). The action in the form \eqref{E-model action} can be derived directly in exactly the same way as above. Explicitly, assuming that $\omega$ has a double pole at infinity, as in \cite{Lacroix:2020flf}, one first fixes the $F$ symmetry by setting the edge mode at infinity equal to the identity. As recalled at the start of this section, this then removes the constant term in both components of the Lax connection. The exact same procedure as above then applies, with the absence of constant terms in the Lax connection reducing \eqref{deg E-model condition} to $\Ec ( \jb_\bz \Lc_\pm) = \pm ( \jb_\bz \Lc_\pm )$ which was the condition used in \cite{Lacroix:2020flf}. In particular, this condition now implies that $\jb_\bz \Lc_\pm \in \va_\pm$.
\end{remark}

\section{Examples} \label{sec: examples}

In \S\ref{sec: review} and \S\ref{sec:reductionto2d} we presented a general construction of integrable degenerate $\Ec$-models from $4$d Chern-Simons theory, with the final $2$d action given in \eqref{E-model action}. In practice, starting from a choice of meromorphic $1$-form $\omega$ one should build the associated defect Lie algebra $\d$, identify its non-degenerate bilinear form $\langle\!\langle \cdot, \cdot \rangle\!\rangle_\d$ and work out the Lie group structure of the defect Lie group $D$. The real vector space of rational functions $R'_{\Pi\bzeta}(\g^\CC)^\Pi$, as defined in \S\ref{sec: vs Rg}, may then be used to explicitly construct the subspaces $\va_\pm$ of $\d$. By making a choice of isotropic Lie subalgebra $\k \subset \d$ one is then able to explicitly construct the projectors $W^\pm_h$ defined by \eqref{projector W degenerate} in terms of which the action \eqref{E-model action} is expressed. In this section we apply this procedure to explicitly construct the pseudo-chiral model and the bi-Yang-Baxter model.

\subsection{Pseudo-chiral model}

The first example we consider is that of a meromorphic $1$-form with a $4^{\rm th}$ order pole at the origin and two simple zeroes at $\pm a$ with $a>0$, namely
\begin{equation*}
\label{ec:omega4pole}
\omega = \frac{a^2-z^2}{z^4} \rd z\,. 
\end{equation*}
The defect Lie algebra \eqref{defect alg} for this choice of meromorphic 1-form is given by
\begin{equation}
\label{ec:dla4pole}
    \dla = \g \otimes \mathbb{R}[\varepsilon]/(\varepsilon^4) \,.
\end{equation}
We denote the elements of $\dla$ by $\ms u^p \coloneqq \ms u \otimes \varepsilon^p$ with $p=0,1,2,3$. The bilinear form \eqref{form on gz rc} on the defect Lie algebra reads
\begin{equation}
\label{ec:pd4 form}
    \langle\!\langle \ms u^p,\ms v^q\rangle \!\rangle_{\dla} =
    \begin{cases}
    \begin{array}{cc}
        a^2\langle \ms u,\ms v\rangle & \text{if} \quad p+q=3 \\
        -\langle \ms u,\ms v\rangle  & \text{if} \quad p+q=1 \\
        0 & \text{otherwise} \,.
    \end{array}
    \end{cases}
\end{equation}

Next, let us describe the Lie group structure on the defect Lie group $D$ with Lie algebra \eqref{ec:dla4pole}. It is given by the $3^{\rm rd}$ order jet bundle $J^3G$ of the Lie group $G$, which in the right trivialisation is isomorphic to $G \times \g \times \g \times \g$. That is, a general element $h \in D \cong G \times \g \times \g \times \g$ can be expressed as a tuple $h=(g,\ms u, \ms v,\ms w)$, with $g\in G$ and $\ms u, \ms v,\ms w\in \g$. The group product and inverse on $D$ are then given by\footnote{The multiplication law and the inverse differ from \cite{Vizman,Lacroix:2020flf} by a normalization convention.} \cite{Vizman}
\begin{align*}
    (g,\ms u, \ms v,\ms w)(\Tilde{g},\ms x,\ms y,\ms z)&=\left(g\Tilde{g}, \ms u +\mathrm{Ad}_g \ms x,\ms v + \mathrm{Ad}_g \ms y+\frac{1}{2}[\ms u, \mathrm{Ad}_g \ms x], \right. \\
    &\qquad \left. \ms w+\mathrm{Ad}_g \ms z + \frac{2}{3}[\ms u, \mathrm{Ad}_g \ms y]+\frac{1}{3}[\ms v, \mathrm{Ad}_g \ms x]+\frac{1}{6}[\ms u,[\ms u, \mathrm{Ad}_g \ms x]]\right),\\
     (g,\ms u, \ms v, \ms w)^{-1}&=\left(g^{-1},-\mathrm{Ad}_g^{-1}\ms u,-\mathrm{Ad}_g^{-1} \ms v,-\mathrm{Ad}_g^{-1} \ms w+\frac{1}{3}\mathrm{Ad}_g^{-1}[\ms u, \ms v]\right) \,.
\end{align*}

To specify the kernels of the operators $W_h^{\pm}$, as in \eqref{projector W degenerate}, we need to make a choice of Lagrangian subalegbra $\k \subset \d$. A natural choice in the present case is
\begin{equation} \label{k choice pseudo-chiral}
\k = \g \otimes \varepsilon^2\mathbb{R}[\varepsilon]/(\varepsilon^4),
\end{equation}
which is easily seen to be Lagrangian. One additional nice feature of this choice of $\k$ is that it is an ideal in $\d$ and thus, for any $h\in D$ we have $\mathrm{Ad}_h^{-1}\k = \k$. Hence, 
\begin{equation}
    \ker W^{\pm}_h= \g \otimes \varepsilon^2\mathbb{R}[\varepsilon]/(\varepsilon^4)=\{\ms y^{2}+\ms z^3 \,|\, \ms y,\ms z \in \g\}\,.
\end{equation}
On the other hand, specifying the image of $W^{\pm}_h$ requires identifying the diagonal subalgebra $\f = \im \Delta$ and the subspaces $\va_\pm$ defined in \eqref{v pm def}. Starting with $\f$, the diagonal embedding for the defect Lie algebra \eqref{ec:dla4pole} is simply $\ms w \mapsto \ms w^0$, so that
\begin{equation}
\label{f 4 order pole}
    \f = \im \Delta = \{ \ms w^0 \,|\, \ms w \in \g\}\,.
\end{equation}
To describe $\va_\pm$ we begin by identifying the spaces of rational functions $R'_{\Pi \bzeta_\pm}(\g^\CC)^\Pi$ corresponding to the meromorphic 1-form \eqref{ec:omega4pole}. By partitioning the set of zeroes $\Pi \bzeta = \bzeta = \{a, -a\}$ of $\omega$ as $\Pi\bzeta_\pm = \bzeta_\pm = \{\pm a\}$, we find
\begin{equation}
    R'_{\Pi \bzeta_\pm}(\g^\CC)^\Pi = \left\{\frac{\ms x}{z\mp a} \, \bigg | \, \ms x\in \g\right\} \,,
\end{equation}
so that expanding such rational functions to $4^{\rm th}$ order at the origin gives
\begin{align}
\label{vplus 4 order pole}
    \va_+ & = \jb_z\left(R'_{\Pi \bzeta_+}(\g^\CC)^\Pi\right)=\left\{-\frac{\ms x^0}{a}-\frac{\ms x^1}{a^2}-\frac{\ms x^2}{a^3}-\frac{\ms x^3}{a^4} \,\,\bigg|\,\, \mathsf{x}\in \g\right\},\\
\label{vminus 4 order pole}    
   \va_- &=\jb_z\left(R'_{\Pi \bzeta_-}(\g^\CC)^\Pi\right)= \left\{\frac{\ms x^0}{a}-\frac{\ms x^1}{a^2}+\frac{\ms x^2}{a^3}-\frac{\ms x^3}{a^4} \,\,\bigg|\,\, \mathsf{x}\in \g\right\}\,. 
\end{align}
Then $\im W^{\pm}_h = \f \oplus\va_\pm$. With the explicit expressions for the image and kernel of the projectors, we may proceed with the computation of $W^{\pm}_h (h^{-1}\partial_\pm h)$. 

In order to simplify the discussion, we start by fixing both the $K$-symmetry and the $F$-symmetry. First, we note that the Lie group $K$ with Lie algebra $\k$ is identified with the subgroup $\{\mathrm{id}\} \times \{0\} \times \g \times \g$ of $G \times \g \times \g \times \g$. On the other hand, the Lie group $F$ with Lie algebra $\f$ is identified with the subgroup $G\times\{0\}\times\{0\}\times\{0\}$. Fixing both of these gauge symmetries, implies that our physical degree of freedom will be described by a representative of the class of $h \in C^\infty(\Sigma, D)$ in the double coset $K \setminus D / F$. By a slight abuse of notation, we will also denote it by $h=(\mathrm{id},\ms u,0,0)$. We then have
\begin{equation}
\label{hdh for 4 pole}
    h^{-1}\rd h = \rd \ms u^1-\frac{1}{2}[\ms u,\rd \ms u]^2+\frac{1}{6}[\ms u,[\ms u,\rd \ms u]]^3 \,.
\end{equation}
In order to find the explicit action of $W^\pm_h$ on $h^{-1}\partial_\pm h$, we decompose the latter with respect to the direct sum decomposition $\dla = \ker W^\pm_h \dotplus \im W^\pm_h$. Focusing first on $h^{-1}\partial_+ h$, we look for $\ms w, \ms x,\ms y,\ms z \in \g$ such that
\begin{equation}
\label{hdh decomposition}
    h^{-1}\partial_+ h = \left(\ms y^2 + \ms z^3\right) + \left(\ms w^0 -\frac{\ms x^0}{a}-\frac{\ms x^1}{a^2}-\frac{\ms x^2}{a^3}-\frac{\ms x^3}{a^4}\right)\,,
\end{equation}
which will then give
\begin{equation*}
    W^+_h (h^{-1}\partial_+ h) = \ms w^0 -\frac{\ms x^0}{a}-\frac{\ms x^1}{a^2}-\frac{\ms x^2}{a^3}-\frac{\ms x^3}{a^4}.
\end{equation*}
Explicitly decomposing the $\sigma^+$-component of \eqref{hdh for 4 pole} as in \eqref{hdh decomposition} we find
\begin{subequations} \label{Wpm explicit pseudo-chiral}
\begin{equation}
    W^+_h (h^{-1}\partial_+ h) = \partial_+ \ms u^1+ \frac{\partial_+\ms u^2}{a}+\frac{\partial_+ \ms u^3}{a^2}\,.
\end{equation}
By a completely analogous argument we get
\begin{equation}
    W^-_h (h^{-1}\partial_- h) =  \partial_- \ms u^1- \frac{\partial_-\ms u^2}{a}+\frac{\partial_- \ms u^3}{a^2}\,.
\end{equation}
\end{subequations}
Using the expression for the bilinear form \eqref{ec:pd4 form} we then obtain
\begin{equation}
    \langle\!\langle W^\pm_h (h^{-1}\partial_\pm h) , h^{-1}\partial_\mp h\rangle\!\rangle = \pm a\langle \partial_+ \ms u,\partial_- \ms u\rangle
\pm \frac{a^2}{2}\langle \ms u, [\partial_+ \ms u, \partial_- \ms u] \rangle.
\end{equation}

Finally, we compute the Wess-Zumino term \eqref{ec:wesszumino}. Since the Cartan $3$-form is cubic in $\widehat{h}^{-1} \rd\widehat{h}$, with $\widehat{h} \in C^\infty(\Sigma \times I, D)$, and the latter has no term of Takiff degree $0$ by \eqref{hdh for 4 pole}, or its analogue for $\widehat{h} = (\rm{id}, \widehat{\ms u}, 0, 0)$, it follows from the explicit form \eqref{ec:pd4 form} of the bilinear form on $\d$ that only the term cubic in $\rd \widehat{\ms u}$ can contribute, so that
\begin{equation}
     I^{\mathrm{WZ}}[h]=-\frac{1}{6}\int_{\Sigma\times I}a^2\langle \rd  \widehat{\ms u},[\rd  \widehat{\ms u},\rd  \widehat{\ms u}]\rangle = \frac{a^2}{6}\int_{\Sigma}\langle \ms u,[\rd \ms u,\rd \ms u]\rangle \,.
\end{equation}

The $2$d action \eqref{E-model action} of the integrable degenerate $\Ec$-model corresponding to the meromorphic 1-form \eqref{ec:omega4pole}, the choice of Lagrangian subalgebra $\k \subset \d$ in \eqref{k choice pseudo-chiral} and the split $\bzeta_\pm = \{\pm a\}$ of the zeroes of $\omega$, is therefore the $\sigma$-model with target space $K \setminus D / F \cong \g$ and action given by
\begin{equation}
\label{ec:pseudochiralaction}
    S[\ms u]=\int_\Sigma \Big( a\langle \partial_+ \ms u,\partial_- \ms u\rangle+\frac{a^2}{3}\langle \ms u,[\partial_+ \ms u,\partial_- \ms u]\rangle \Big) \rd \sigma^+ \wedge \rd \sigma^-
\end{equation}
for $\ms u \in C^\infty(\Sigma, \g)$, which is the pseudo-chiral model of Zakharov and Mikhailov \cite{Zakharov1978}. 

\subsubsection{Lax connection}

Having found the action of the $2$d integrable field theory, we now proceed with the computation of its Lax connection. Its light-cone components are given by \eqref{Lax connection E-model} where $\bm p : \f^\perp \SimTo R_{\Pi \bzeta}(\g^\CC)^\Pi$ is the inverse of $\jb_z$ and
\begin{equation} \label{f perp decomp}
\f^\perp = \f\oplus \va_+ \oplus \va_-
\end{equation}
with $\f$, $\va_+$ and $\va_-$ given in \eqref{f 4 order pole}, \eqref{vplus 4 order pole} and \eqref{vminus 4 order pole}, respectively. Hence, the action of $\bm p$ on an element of $\f^\perp$ decomposed with respect to \eqref{f perp decomp} is simply
\begin{equation}
    \bm p \left(\ms w^0 -\frac{\ms x^0}{a}-\frac{\ms x^1}{a^2}-\frac{\ms x^3}{a^3}-\frac{\ms x^3}{a^4}+\frac{\ms y^0}{a}-\frac{\ms y^1}{a^2}+\frac{\ms y^3}{a^3}-\frac{\ms x^3}{a^4}\right) = \ms w + \frac{\ms x}{z-a}+ \frac{\ms y}{z+a}\,.
\end{equation}
The action of $\bm p$ on $W^\pm_{h}(h^{-1}\partial_\pm h)$ given in \eqref{Wpm explicit pseudo-chiral} can now be computed to give the light-cone components of the Lax connection \eqref{Lax connection E-model}, namely we find
\begin{equation}
    \Lc_\pm = \frac{-a^2}{z\mp a}\partial_\pm \ms u  \mp a \partial_\pm \ms u \,.
\end{equation}
The zero curvature equation for this Lax connection is equivalent to 
\begin{equation}
    \partial_+\partial_- \ms u - \frac{a}{2}[\partial_+ \ms u,\partial_- \ms u]=0\,,
\end{equation}
which corresponds to the equation of motion of $\ms u$ for the action \eqref{ec:pseudochiralaction}, thus proving the Lax integrability of the model.

\subsection{Bi-Yang-Baxter \texorpdfstring{$\sigma$}{s}-model} 

The second example we consider is the bi-Yang-Baxter $\sigma$-model \cite{KlimcikBi,Klimcik:2014bta}. Following the conventions used in \cite{Delduc:2015xdm,Delduc:2019whp}, we take the meromorphic 1-form
\begin{equation}
\label{byb model omega}
    \omega=\frac{16 K z}{\zeta^2(z-z_+)(z-z_-)(z-\tilde{z}_+)(z-\tilde{z}_-)} \rd z
\end{equation}
where $K\in \mathbb{R}$. The four simple poles $z_\pm, \tilde{z}_\pm \in \CC$ and the coefficient $\zeta \in \mathbb{R}$ are related to the two real deformation parameters $\eta$ and $\tilde{\eta}$ of the model by
\begin{gather} \label{ec:zpluszminus}
    z_\pm =\frac{-2\rho \pm i \eta}{\zeta}\,, \quad \tilde{z}_\pm=-\frac{2+2\rho \pm i \tilde{\eta}}{\zeta}\,, \quad \rho=-\tfrac{1}{2}\left(1-\frac{\eta^2-\tilde{\eta}^2}{4}\right),\\
    \zeta^2=\left(1+\frac{(\eta+\tilde{\eta})^2}{4}\right)\left(1+\frac{(\eta-\tilde{\eta})^2}{4}\right)\,.
\end{gather}
Note also that $\omega$ has two simple zeroes, at $0$ and $\infty$. The defect Lie algebra \eqref{defect alg} for this choice of meromorphic 1-form is given by
\begin{equation}
\label{byb defect lie algebra}
    \dla = \big(\g^{\mathbb{C}}\otimes \mathbb{C}[\varepsilon]/(\varepsilon)\big) \times \big(\g^{\mathbb{C}}\otimes \mathbb{C}[\Tilde{\varepsilon}]/(\Tilde{\varepsilon})\big) \cong \g^\CC \times \g^\CC,
\end{equation}
where we recall that each factor is treated as a real vector space.
Therefore elements of $\dla$ are given by tuples $(\ms u,\ms v)\in \dla$ with $\mathsf{u},\mathsf{v}\in \g^{\mathbb{C}}$.
The bilinear form \eqref{form on gz 1} on the defect Lie algebra reads
\begin{equation}
\label{ec:byb form}
    \langle\!\langle (\mathsf{u},\mathsf{v}),(\tilde{\mathsf{u}},\tilde{\mathsf{v}})\rangle \!\rangle_{\dla} =\tfrac{4K}{\eta}\Im\langle \mathsf{u},\tilde{\mathsf{u}}\rangle +\tfrac{4K}{\tilde{\eta}}\Im\langle \mathsf{v},\tilde{\mathsf{v}}\rangle.
\end{equation}
The defect Lie group $D$ with Lie algebra $\d$ is simply $G^{\CC}\times G^{\CC}$, a general element of which is a tuple $(h,\Tilde{h})$ with $h,\tilde{h} \in G^{\CC}$.

To specify the kernels of the projection operators defined by \eqref{projector W degenerate} we need to choose a Lagrangian subalgebra $\k \subset \d$. Following \cite{Delduc:2019whp}, we take two skew-symmetric solutions $R,\tilde{R}\in \mathrm{End}\,\g$ to the modified Yang-Baxter equation with $c=\ms i$ in terms of which we define the Lie subalgebra
\begin{equation}
\label{ec:byblagsubalgebra}
    \k \coloneqq \g_R\times \g_{\tilde{R}}=\{((R-\ms i)\mathsf{x},(\tilde{R}-\ms i)\mathsf{y})\,|\, \mathsf{x},\mathsf{y}\in \g\}
\end{equation}
which is seen to be Lagrangian with respect to the bilinear form \eqref{ec:byb form}. To simplify the discussion we will gauge fix the $K$-symmetry. Let $K= G_R \times G_{\tilde{R}} \subset G^{\CC}\times G^{\CC}$ be the Lie group with Lie algebra $\k = \g_R\times \g_{\tilde{R}}$. Following \cite{Delduc:2019whp}, we assume that the direct sum decomposition $\g^{\CC} \times \g^{\CC}= \k \dotplus (\g \times \g)$ lifts to the group level, that is, $G^{\CC}\times G^{\CC} = K (G \times G)$ so that a natural parametrisation of the quotient $K \setminus (G^\CC \times G^\CC)$ is then given by $G\times G$. In this way, our physical degrees of freedom will be described by a representative of the class of $h \in C^\infty(\Sigma, G^\CC \times G^\CC)$ in the coset $K \setminus (G^\CC \times G^\CC)$ which we denote by $(g,\tilde{g})$ with $g,\tilde{g} \in C^\infty(\Sigma, G)$. Hence, from \eqref{projector W degenerate} we have
\begin{equation}
    \mathrm{ker}\, W^\pm_{(g,\Tilde{g})} = \mathrm{Ad}_{(g,\Tilde{g})}^{-1}\g_R \times \g_{\Tilde{R}}=\left\{\left((R_g-\ms i)\mathrm{Ad}_g^{-1}\mathsf{x},(\tilde{R}_{\tilde{g}}-\ms i)\mathrm{Ad}_{\tilde{g}}^{-1}\mathsf{y}\right)\,|\, \mathsf{x},\mathsf{y}\in \g\right\}\,,
\end{equation}
where we have defined $R_g = \mathrm{Ad}_g^{-1} \circ R \circ \mathrm{Ad}_g$, and similarly for $\Tilde{R}_{\Tilde{g}}$.

On the other hand, the images of $W^{\pm}_{(g,\Tilde{g})}$ are given in terms of the subalgebra $\f = \im \Delta$ and the subspaces $\va_\pm$ defined in \eqref{v pm def}. The diagonal embedding for the defect Lie algebra \eqref{byb defect lie algebra} is simply $\ms a \mapsto \ms (\ms a,\ms a)$, so that
\begin{equation}
\label{f byb model}
    \f =\{(\ms a,\ms a) \,|\, \mathsf{a}\in \g\}\,.
\end{equation}
To determine $\va_\pm$ we must first identify the space of rational functions $R'_{\Pi \bzeta_\pm}(\g^\CC)^\Pi$ corresponding to the meromorphic 1-form \eqref{byb model omega}. Fixing the partition of the set of zeroes $\Pi \bzeta = \bzeta = \{0, \infty\}$ of $\omega$ to be $\Pi\bzeta_+ = \bzeta_+ = \{\infty\}$ and $\Pi\bzeta_- = \bzeta_- = \{0\}$ we have
\begin{equation}
    R'_{\Pi \bzeta_+}(\g^\CC)^\Pi = \left\{\ms b z \,  | \, \ms b\in \g\right\} \,,\quad R'_{\Pi \bzeta_-}(\g^\CC)^\Pi = \left\{\frac{\ms b}{z} \,   | \, \ms b\in \g\right\}.
\end{equation}
Expanding such rational functions at the set of independent poles $\bz = \{ z_+, \tilde{z}_+ \}$ of $\omega$ yields
\begin{align}
\label{vplus byb model}
    \va_+ & = \jb_z\left(R'_{\Pi \bzeta_+}(\g^\CC)^\Pi\right)=\left\{\left(\ms b z_+,\ms b\tilde{z}_+\right)\,|\, \ms b\in \g\right\}, \\
\label{vminus byb model}    
   \va_- &=\jb_z\left(R'_{\Pi \bzeta_-}(\g^\CC)^\Pi\right)= \left\{\left(\frac{\ms b}{z_+},\frac{\ms b}{\tilde{z}_+}\right)\,|\, \ms b\in \g \right\}\,.
\end{align}
We then have $\im W^{\pm}_{(g,\Tilde{g})} = \f \oplus\va_\pm$ and we may now proceed with the computation of $W^{\pm}_{(g,\Tilde{g})}(j_\pm,\Tilde{\jmath}_\pm)$ where we defined $j_{\pm}\coloneqq g^{-1}\partial_\pm g$ and $\Tilde{\jmath}_\pm \coloneqq \Tilde{g}^{-1}\partial_\pm \Tilde{g}$. 

In order to find the explicit action of $W^\pm_{(g,\Tilde{g})}$ on $(j_\pm,\Tilde{\jmath}_\pm)$, we decompose the latter with respect to the direct sum decomposition $\dla = \ker W^\pm_{(g,\Tilde{g})} \dotplus \im W^\pm_{(g,\Tilde{g})}$. Explicitly, focusing first on $(j_+, \tilde{\jmath}_+)$, we look for $\ms a, \ms b, \ms x, \ms y \in \g$ such that 
 \begin{equation}
 \label{ec:uv}
     \left(j_+,\Tilde{\jmath}_+\right)=(\ms a, \ms a)+\left(\ms b z_+,\ms b \Tilde{z}_+\right)+\left((R_g-\ms i)\mathrm{Ad}_g^{-1}\mathsf{x},(\tilde{R}_{\Tilde{g}}-\ms i)\mathrm{Ad}_{\Tilde{g}}^{-1}\mathsf{y}\right).
\end{equation}
To match the notation from \cite{Delduc:2019whp} it is convenient to introduce
\begin{equation}
    J_\pm = \frac{1}{1\pm\frac{\eta}{2} R_g \pm\frac{\Tilde{\eta}}{2}R_{\Tilde{g}}}(j_\pm - \Tilde{\jmath}_\pm)\,,
\end{equation}
in terms of which the solution to \eqref{ec:uv} can be conveniently written as
\begin{gather*}
    \ms a = j_+ + \bigg( \rho - \frac{\eta}{2} R_g \bigg) J_+ = \tilde{\jmath}_+ + \bigg( 1+\rho + \frac{\tilde{\eta}}{2} \tilde{R}_{\tilde{g}} \bigg) J_+, \\
    \ms b = \frac{\zeta}{2} J_+,
    \qquad \mathrm{Ad}_{g}^{-1} \ms x = \frac{\eta}{2} J_+,
    \qquad \mathrm{Ad}_{\Tilde{g}}^{-1} \ms y  = -\frac{\tilde{\eta}}{2} J_+\,.
\end{gather*}
In particular, given that $W^+_{(g,\Tilde{g})}$ is a projector, its action on \eqref{ec:uv} is given by the first two terms on the right hand side, namely we have
\begin{equation}
\label{ec:wpbybm}
    W_{(g,\Tilde{g})}^+ (j_+, \Tilde{\jmath}_+) 
    = \bigg( j_+ - \frac{\eta}{2} (R_g - \ii) J_+, \tilde{\jmath}_+ + \frac{\tilde{\eta}}{2} (\tilde{R}_{\tilde{g}} - \ii) J_+ \bigg) \,.
\end{equation}
Similarly, to compute the action of $W^-_{(g,\Tilde{g})}$ on $(j_-,\Tilde{\jmath}_-)$ we look again for $\ms a, \ms b, \ms x, \ms y \in \g$ but this time such that 
 \begin{equation}
 \label{ec:uv2}
     (j_-,\Tilde{\jmath}_-)=(\ms a, \ms a)+\left(\frac{\ms b}{z_+},\frac{\ms b}{\Tilde{z}_+}\right)+\left((R_g-\ms i)\mathrm{Ad}_g^{-1}\mathsf{x},(\tilde{R}_{\Tilde{g}}-\ms i)\mathrm{Ad}_{\Tilde{g}}^{-1}\mathsf{y}\right) \,.
 \end{equation}
Doing so, we find
\begin{equation}
\label{ec:wmbybm}
    W_{(g,\Tilde{g})}^- (j_-,\Tilde{\jmath}_-)
    = \bigg( j_- + \frac{\eta}{2} (R_g - \ii) J_-, \tilde{\jmath}_- - \frac{\tilde{\eta}}{2} (\tilde{R}_{\tilde{g}} - \ii) J_- \bigg)\,.
\end{equation}
Using the expression for the bilinear form \eqref{ec:byb form} we then obtain
\begin{equation}
    \langle\!\langle  W_{(g,\Tilde{g})}^\pm (j_\pm,\Tilde{\jmath}_\pm ), (j_\mp,\Tilde{\jmath}_\mp)\rangle \!\rangle_{\d} = \pm 2K \langle j_+ - \tilde{\jmath}_+, J_- \rangle
\end{equation}
where we used the fact that $\langle j_- - \tilde{\jmath}_-, J_+ \rangle = \langle j_+ - \tilde{\jmath}_+, J_- \rangle$ which follows from the skew-symmetry of $R$ and $\tilde R$.

On the other hand, it is immediate to verify that the Wess-Zumino term vanishes identically. We thus find that the $2$d action \eqref{E-model action} is given by
\begin{equation}
    S[g,\Tilde{g}]=K\int_{\Sigma}\left\langle j_+-\Tilde{\jmath}_+,J_-\right\rangle \rd \sigma \wedge\rd \tau,
\end{equation}
matching the action of the bi-Yang-Baxter $\sigma$-model as written in \cite[(2.2)]{Delduc:2015xdm}. 

\subsubsection{Lax connection}

The Lax connection is given by \eqref{Lax connection E-model}, which for this specific example becomes 
\begin{equation}
\Lc_\pm\big( (g,\Tilde{g}) \big) = \bm p \big( W^\pm_{(g,\Tilde{g})}(j_\pm,\Tilde{\jmath}_\pm) \big)\,,
\end{equation}
where $\bm p : \f^\perp \SimTo R_{\Pi \bzeta}(\g^\CC)^\Pi$ is the inverse of $\jb_\bz$ with
\begin{equation} \label{f perp biYB}
\f^\perp = \f\oplus \va_+ \oplus \va_-
\end{equation}
with $\f$, $\va_+$ and $\va_-$ given in \eqref{f byb model}, \eqref{vplus byb model} and \eqref{vminus byb model}, respectively, so that the action of $\bm p$ on an element in $\f^\perp$ decompose with respect to \eqref{f perp biYB} is simply
\begin{equation}
    \bm p \left((\ms a,\ms a)+(\ms b z_+,\ms b \Tilde{z}_+)+\left(\frac{\ms c}{z_+},\frac{\ms c}{\Tilde{z}_+}\right)\right) = \ms a + \ms b z + \frac{\ms c}{z}\,.
\end{equation}
Therefore, decomposing $W^\pm_{(g,\Tilde{g})}(j_\pm,\Tilde{\jmath}_\pm)$ with respect to \eqref{f perp biYB} we find 
\begin{equation}
   \Lc_+ = B_++\frac{\zeta}{2}z J_+   \,,\quad \Lc_-=B_-+\frac{\zeta}{2}z^{-1}J_-
\end{equation}
where we have defined 
\begin{equation}
    B_{\pm}= j_\pm + \bigg( \rho \mp \frac{\eta}{2}R_g \bigg) J_\pm \,,
\end{equation}
with $\rho$ defined in \eqref{ec:zpluszminus}. The expressions for the components of the Lax connection coincide, up to a conventional sign, with \cite[(2.18)]{Delduc:2015xdm}.

\section{Outlook}

\def\Ann{\raisebox{-.2mm}{\begin{tikzpicture}[scale=0.065]
  \path [draw=none,fill=gray, fill opacity = 0.1] (0,-1) circle (2);
  \path [draw=none,fill=white, fill opacity = 1] (0,-1) circle (.7);
  \draw[solid] (0,-1) circle (.7);
  \draw[solid] (0,-1) circle (2);
\end{tikzpicture}}}
\def\Disc{\raisebox{-.2mm}{\begin{tikzpicture}[scale=0.065]
  \path [draw=none,fill=gray, fill opacity = 0.1] (0,-1) circle (2);
  \draw[solid] (0,-1) circle (2);
\end{tikzpicture}}}
\def\sDisc{\raisebox{-.2mm}{\begin{tikzpicture}[scale=0.05]
  \path [draw=none,fill=gray, fill opacity = 0.1] (0,-1) circle (2);
  \draw[solid] (0,-1) circle (2);
\end{tikzpicture}}}

In this work we have shown that the $4$d Chern-Simons action introduced by Costello and Yamazaki in \cite{Costello:2019tri} with the most general meromorphic $1$-form $\omega$ gives rise, when passing to $2$d following the approach of \cite{Benini:2020skc, Lacroix:2020flf}, to the actions of integrable versions of degenerate $\Ec$-models, or dressing cosets, introduced by Klim\v{c}\'\i{}k and \v{S}evera \cite{Klimcik:1995dy}.

Notably, this article resolves one of the open problems mentioned in \cite{Lacroix:2020flf}, namely removing the assumption that the meromorphic 1-form $\omega$ should have a double pole at infinity, thus allowing it to be completely arbitrary. There are, however, a number of other interesting problems that remain open which apply to the degenerate setting as well. We summarize them here for completeness. 

\medskip

The first is related to the Hamiltonian description of the integrable degenerate $\Ec$-models constructed in the present work. Indeed, as recalled in the introduction, establishing the complete integrability of a $2$d field theory requires moving to the Hamiltonian formalism and showing that the Poisson bracket of the spatial component of the Lax connection with itself takes the non-ultralocal $r/s$-form \cite{Maillet:1985fn, Maillet:1985ek} with twist function. This, in turn, is equivalent to recasting the $2$d field theory in question as a classical dihedral affine Gaudin model, see \cite{Vicedo:2017cge} and also \cite{Delduc:2019bcl,Lacroix:2019xeh}. And although we have not shown this explicitly here, it follows indirectly from \cite{Vicedo:2019dej} where a Hamiltonian analysis of $4$d Chern-Simons theory was performed and it was shown that the Poisson bracket of $\Lc_\sigma$, i.e. the spatial component of the gauge field in the gauge $A_{\bar z} = 0$, with itself is precisely of the required form with the twist function $\varphi(z)$ determined by the meromorphic $1$-form $\omega = \varphi(z) \rd z$.

It would, nevertheless, be interesting to perform the Hamiltonian analysis of the integrable degenerate $\Ec$-models constructed here to directly show that the spatial component of their Lax connection has a Poisson bracket with itself of the expected $r/s$-form with twist function determined by $\omega$. In particular, sufficient conditions on the $\Ec$-model data ensuring its integrability in the Hamiltonian sense were given in \cite{Klimcik:2021bjy}. These are anaolgous to the sufficient conditions on the $\Ec$-model data given in \cite{Severa:2017kcs}, see also \cite{Lacroix:2020flf} and Lemma \ref{lem: p property} above, ensuring the existence of a Lax connection. It would therefore be interesting to check that the integrable degenerate $\Ec$-models constructed from $4$d Chern-Simons satisfy the sufficient conditions of \cite{Klimcik:2021bjy}.

\medskip

The second interesting open direction is to determine the relationship between 4d Chern-Simons and the usual 3d Chern-Simons theory. Indeed, it was shown in \cite{Severa:2016prq} (see also \cite{Arvanitakis:2022bnr}) that the non-degenerate $\Ec$-model on $S^1\times \mathbb{R}$ can be obtained from $3$d Chern–Simons theory for the Lie group $D$ on $\Disc \times \mathbb{R}$, with $\Disc$ a disc, by imposing twisted self-dual boundary conditions on the gauge field $A$ of the form
\begin{equation} \label{twisted SD bc}
\ast A|_{\partial (\sDisc \times \RR)} = \Ec A|_{\partial (\sDisc \times \RR)}
\end{equation}
on the boundary $\partial \, (\Disc \times \RR) \cong S^1 \times \RR$. It was further shown in \cite{Severa:2016prq} that the $\sigma$-model on $K \setminus D$ can be obtained from $3$d Chern–Simons theory on a hollowed out cylinder $\Ann \times \mathbb{R}$, with $\Ann$ an annulus, by imposing twisted self-dual boundary conditions on the gauge field as in \eqref{twisted SD bc} at the outer boundary and imposing at the inner boundary $\Sigma_{\mathrm{inn}}$ a condition of the form
\begin{equation} \label{inner bdry bc}
A|_{\Sigma_{\mathrm{inn}}}\in \Omega^{1}(\Sigma_{\mathrm{inn}},\k),
\end{equation}
with $\k \subset \d$ a Lagrangian subalgebra. Given the similarities between the boundary conditions considered in the present $4$d Chern-Simons context, in particular \eqref{non-deg E-model condition} as in \cite{Lacroix:2020flf} or \eqref{deg E-model condition} as considered here together with \eqref{constraint L h}, and the boundary conditions \eqref{twisted SD bc} and \eqref{inner bdry bc} considered in the $3$d Chern-Simons setting, it would be interesting to understanding whether there is any deeper connection between these two theories.

\medskip

Finally, there is at least one other interesting direction in which to generalise the whole construction, which is to try and describe from a $4$d Chern-Simons perspective the class of $2$d integrable field theories whose Lax connections are equivariant under the action of a cyclic group $\ZZ_T$ for some $T \in \ZZ_{\geq 2}$. Such $2$d integrable field theories should, on general grounds, arise from $4$d Chern-Simons theory on a certain orbifold quotient of $\Sigma \times \CP$ by $\ZZ_T$. That is, one should start from $4$d Chern-Simons theory on $\Sigma \times \CP$ but impose that the gauge field $A$ be equivariant with respect to actions of the cyclic group on $\CP$ and on $\g$. Such a setting has already been considered in \cite{Schmidtt:2020dbf} in relation to a specific $2$d integrable field theory, namely the $\lambda$-model. More generally, it would be interesting to extend the results of \cite{Benini:2020skc} to the equivariant setting and use this to construct integrable (non-)degenerate $\Ec$-models with equivariant Lax connections along the lines of \cite{Lacroix:2020flf} and the present paper for general $\omega$.

\subsection*{Acknowledements}

B.V. would like to thank S. Lacroix for fruitful discussions in relation to degenerate $\Ec$-models. J.L. would like to thank L. Cole and R. Cullinan for great discussions on 4d CS theory. J.L. would also like to thank H. Falomir and P. Pisani for their endless support as PhD advisors. B.V. gratefully acknowledges the support of the Leverhulme Trust through a Leverhulme Research Project Grant (RPG-2021-154). J.L. gratefully acknowledges the support of CONICET through a PhD scholarship. 

\printbibliography[title=\textsc{References}]

%\bibliography{literature}

\end{document}